\RequirePackage{amsmath}
\documentclass[a4paper, 11pt]{article}

\usepackage[margin=1.15in]{geometry}
\usepackage[utf8]{inputenc} 
\usepackage{amsfonts}
\usepackage{amssymb}
\usepackage{amsthm}

\newtheorem{theorem}{Theorem}
\newtheorem{proposition}[theorem]{Proposition}
\newtheorem{lemma}[theorem]{Lemma}
\newtheorem{corollary}[theorem]{Corollary}

\usepackage{hyperref}

\usepackage{tikz}
\usetikzlibrary{arrows.meta}
\usetikzlibrary{calc}
\usetikzlibrary{positioning}
\usetikzlibrary{math}

\def\2opt{2-Opt heuristic}
\def\kopt{$k$-Opt heuristic}

\def\blue#1{#1}

\title{The Approximation Ratio of the $k$-Opt~Heuristic for the Euclidean Traveling Salesman Problem}
\author{Ulrich A.\ Brodowsky\thanks{Pontsheide 20, 52076 Aachen, Germany ({ulrich.brodowsky@gmx.net}).} 
\and Stefan Hougardy\thanks{Research Institute for Discrete Mathematics and Hausdorff Center for Mathematics, University of Bonn, Lenn\'estr.~2, 53113 Bonn, Germany ({hougardy@dm.uni-bonn.de}, {zhong@uni-bonn.de})
{funded by the Deutsche Forschungsgemeinschaft (DFG, German Research Foundation) under Germany's Excellence Strategy -- EXC-2047/1 -- 390685813.}
}
\and Xianghui Zhong\footnotemark[2]
}

\begin{document}

\maketitle

\begin{abstract}
The $k$-Opt heuristic is a simple improvement heuristic for the Traveling Salesman Problem.
It starts with an arbitrary tour and then repeatedly replaces $k$ edges of the tour  
by $k$ other edges, as long as this yields a shorter tour. 
We will prove that for 2-dimensional Euclidean Traveling Salesman Problems with $n$ cities 
the approximation ratio of the $k$-Opt heuristic is $\Theta(\log n / \log \log n)$. 
This improves the upper bound of $O(\log n)$ given by Chandra, Karloff, and Tovey in 1999~\cite{CKT1999}
and provides for the first time a non-trivial lower bound for the case $k\ge 3$. Our results not only hold
for the Euclidean norm but extend to arbitrary $p$-norms \blue{with $1 \le p < \infty$}. 
\end{abstract}



\section{Introduction}

The Traveling Salesman Problem (TSP) is one of the best studied problems in combinatorial optimization.
Given $n$ cities and their pairwise distances, the task is to find a shortest tour that visits each city exactly once.
This problem is NP-hard~\cite{GJ1979} and it is even hard to approximate to a factor that is polynomial in $n$~\cite{SG1976}.

In the \emph{$d$-dimensional Euclidean TSP}, the cities are points in $\mathbb{R}^d$ and the distance function is the Euclidean distance between the points. From a practical point of view the 2-dimensional Euclidean TSP is of particular importance. 
As most of our results concern the 2-dimensional case we follow a usual convention (see e.g.~\cite{Rei1994}) and  
name the 2-dimensional Euclidean TSP simply \emph{Euclidean TSP}. 
\blue{The Euclidean TSP is also NP-hard~\cite{Pap1977} but it allows a polynomial time approximation scheme~\cite{Aro1998,Mit1999}.}

\blue{Traveling Salesman Problems often appear in practice and they are usually solved using some heuristics.}
One of the simplest of these heuristics is the \emph{\kopt}. 
It starts with an arbitrary tour and then repeatedly replaces $k$ edges of the tour  
by $k$ other edges, as long as this yields a shorter tour. The \kopt\ stops when no further improvement can be made this way. 
A tour that the \kopt\ cannot improve is called \emph{$k$-optimal}.

On real-world instances the \kopt\ achieves surprisingly good results even in the case $k=2$ (see e.g.\ Bentley~\cite{Ben1992}). 
Despite its simplicity the exact approximation ratio of the \kopt\ for Euclidean TSP was not known so far. 
For the special case $k=2$ Chandra, Karloff, and Tovey~\cite{CKT1999} proved in 1999
a lower bound of $\Omega(\frac{\log n}{\log \log n})$ and an upper bound of $O(\log n)$.
For arbitrary (but constant) $k$ the approximation ratio of the \kopt\ for Euclidean TSP was far open: 
no non-trivial lower bound was known and only the upper bound of $O(\log n)$ from the case $k=2$ was known~\cite{CKT1999}.

Our main result closes this gap by determining the approximation ratio of the \kopt\ up to a constant factor: 

\begin{theorem} For constant $k \ge 2$ the 
approximation ratio of the $k$-Opt heuristic for Euclidean TSP instances with $n$ points is $\Theta(\log n / \log \log n)$.
\label{thm:combinedresult}
\end{theorem}

We prove this result by first presenting a new upper bound for the case $k=2$: 

\begin{theorem}
The approximation ratio of the \2opt\ for Euclidean TSP instances with $n$ points is $O(\log n / \log \log n)$.
\label{thm:main}
\end{theorem}

In a second step we are able to extend the known lower bound for the \2opt~\cite{CKT1999}
to the \kopt:

\begin{theorem} For constant $k \ge 2$ the 
approximation ratio of the $k$-Opt heuristic for Euclidean TSP instances with $n$ points is $\Omega(\log n / \log \log n)$.
\label{thm:k-Opt}
\end{theorem}

As we will see in Section~\ref{sec:lower-bound} this lower bound also holds in the case of arbitrary $p$-norms instead of the Euclidean norm 
(see Theorem~\ref{thm:LowerBoundLp}). Clearly, these lower bounds for $\mathbb{R}^2$ also hold for the $d$-dimensional case for $d > 2$.

We will prove Theorem~\ref{thm:main} in \blue{Sections~\ref{sec:notation} to~\ref{sec:proof}} and Theorem~\ref{thm:k-Opt} in Section~\ref{sec:lower-bound}.
Theorem~\ref{thm:combinedresult} is a direct consequence of Theorem~\ref{thm:main} and Theorem~\ref{thm:k-Opt}:
By definition the \kopt\ for $k\ge 2$ always returns a 2-optimal solution. Therefore, 
Theorem~\ref{thm:main} implies that the upper bound for the approximation ratio of the $k$-Opt heuristic is $O(\log n / \log \log n)$.
The lower bound for the \kopt\ follows from Theorem~\ref{thm:k-Opt}.

\subparagraph*{Related Results.} For constant $k\ge 2$ one can decide in polynomial time whether a given tour can be shortened by
replacing at most $k$ edges of the tour by $k$ other edges. 
On real-world Euclidean TSP instances it has been observed that the  2-Opt heuristic 
needs a sub-quadratic number of iterations until it reaches a local optimum~\cite{Ben1992}. However, 
there exist worst-case Euclidean TSP instances for which the 2-Opt heuristic may need an exponential number of 
iterations~\cite{ERV2014}.

For $n$ points embedded into the $d$-dimensional Euclidean space $\mathbb{R}^d$ for some constant $d>2$ the approximation
ratio of the \kopt\ is bounded by $O(\log n)$ from above~\cite{CKT1999} and by  $\Omega(\log n / \log \log n)$
from below by Theorem~\ref{thm:k-Opt}.  

The Euclidean TSP is a special case of the \emph{metric TSP}, i.e., the Traveling Salesman Problem
where the distance function satisfies the triangle inequality. 
The well-known algorithm of Christofides~\cite{Chr2022b} and Serdjukov~\cite{Ser1978} achieves an approximation ratio of $3/2$
for the metric TSP. \blue{The algorithm of} Karlin, Klein, and Oveis Gharan~\cite{KKO2022} 
slightly improves on this. For the metric TSP the 2-Opt heuristic 
has approximation ratio exactly $\sqrt{n/2}$~\cite{HZZ2020}. For constant $k > 2$ 
a lower bound of $\Omega(n^{\frac{2}{3k-3}})$ and 
an upper bound of $O(n^{\frac{1}{k}})$ on the approximation ratio of the $k$-Opt heuristic for metric TSP are known~\cite{Zho2020b}. 
The upper bound of $O(n^{\frac{1}{k}})$ implies that for $k = \Omega(\log n)$ the approximation ratio of the \kopt\ for metric TSP
and therefore also for the Euclidean TSP is constant.
In the special cases $k=3,4,6$ the approximation ratio of the $k$-Opt heuristic for metric TSP is
$\Theta(n^{\frac{1}{k}})$~\cite{{Zho2020b}}.

A very special case of the metric TSP is the 1-2-TSP. In this version all edge lengths
have to be~1 or~2. For the 1-2-TSP the approximation ratio of the 2-Opt heuristic is $3/2$~\cite{KMSV1998}.
The $3$-Opt heuristic for 1-2-TSP has approximation ratio $11/8$~\cite{Zho2021}. For constant $k>3$ 
the approximation ratio of the $k$-Opt heuristic for 1-2-TSP lies between $11/10$ and $11/8$~\cite{Zho2020}.

\subparagraph*{Organization of the paper.}
To prove Theorem~\ref{thm:main}, i.e.\ the upper bound of the 2-Opt heuristic for Euclidean TSP we proceed as follows. 
First we will present in Section~\ref{sec:notation} some properties of Euclidean 2-optimal tours. 
In Section~\ref{sec:uncrossing} we will \blue{reduce} Theorem~\ref{thm:main} to the special case 
where no intersections between the edges of an optimal tour and the edges of a 2-optimal tour exist. 
In this special case we will show that we can partition the edge set of a 2-optimal tour into five sets 
\blue{such that the edges in each of these sets can be oriented in a way that relates to the orientation of}
an optimal tour. The main step then is to prove that for each of these five sets we can bound the total edge length by 
$O(\log n / \log \log n)$ times the length of an optimal tour. 
To achieve this we will relate optimal tours and subsets of the edge set of a 2-optimal tour to some weighted arborescences. 
This relation is studied in Section~\ref{sec:proofidea}.
For weighted arborescences we will provide in Section~\ref{sec:arborescence-lemmas} some bounds for the edge weights. 
These results then will allow us in Section~\ref{sec:proof} to finish the proof of Theorem~\ref{thm:main}.

In Section~\ref{sec:lower-bound} we will prove Theorem~\ref{thm:k-Opt}, i.e.\ the lower bound for the \kopt\ for
Euclidean TSP. 
For this we will modify a construction of instances given in~\cite{CKT1999} for the 2-Opt heuristic to the \kopt\ and to arbitrary $p$-norms
\blue{with $1\le p < \infty$}. 

All but one step in our proof of Theorem~\ref{thm:main} work in arbitrary dimensions. In Section~\ref{sec:higherDimension}
we will provide a 3-dimensional example showing that our current proof \blue{of Theorem~\ref{thm:main}} does not allow to extend \blue{Theorem~\ref{thm:main}}
to higher dimensions. Finally, in Section~\ref{sec:LpNorms} we discuss the extension of \blue{Theorem~\ref{thm:combinedresult}} to
arbitrary $p$-norms \blue{(with $1\le p < \infty$)} and state the most general result of our paper in Theorem~\ref{thm:LpNormResult}.

\section{Euclidean TSP and 2-Optimal Tours}
\label{sec:notation}

An instance of the Euclidean TSP is a finite subset $V\subset \mathbb{R}^2$. 
The task is to find a polygon of shortest total edge length that contains all points of $V$.
Note that by our definition a Euclidean TSP instance cannot contain the same point multiple times.
In the following we will denote the cardinality of $V$ by $n$.

For our purpose it is often more convenient to state the Euclidean Traveling Salesman Problem as a problem on graphs. 
For a given point set $V$ of a Euclidean TSP instance we take a complete graph on the vertex set $V$, i.e., the graph $G=(V,E)$
where $E$ is the set of all $\frac12  n(n-1)$ possible edges on $V$. We assign the Euclidean  
distance between the vertices in $G$ by a function  $c:E(G) \to \mathbb{R}_{> 0}$.
A \emph{tour} in $G$ is a cycle that contains all the vertices of $G$.
The \emph{length} of a tour $T$ in $G$ is defined as $c(T) := \sum_{e\in E(T)} c(e)$.  
An \emph{optimal tour} is a tour of minimum length among the tours in $G$. 
Thus we can restate the Euclidean TSP as a problem in graphs: 
Given a complete graph $G=(V,E)$ on a point set $V\subset \mathbb{R}^2$ and a Euclidean distance function $c:E(G) \to \mathbb{R}_{> 0}$, 
find an optimal tour in $G$. 
Throughout this paper we will use the geometric definition of the Euclidean TSP and the graph-theoretic version of the
Euclidean TSP simultaneously. Thus, a tour for a Euclidean TSP instance $V\subseteq \mathbb{R}^2$ can be viewed as a polygon in  $\mathbb{R}^2$ as well as
a cycle in a complete graph on the vertex set $V$ with Euclidean distance function.

Let $c:E(G)\to\mathbb{R}_{>0}$ be a weight function for the edges of some graph $G=(V,E)$. 
To simplify notation, we will denote the weight of an edge $\{x,y\} \in E(G)$ simply by $c(x,y)$ instead 
of the more cumbersome notation $c(\{x,y\})$. For subsets $F\subseteq E(G)$ we define
$c(F) := \sum_{e\in F} c(e)$. We extend this definition to subgraphs $H$ of $G$ by setting $c(H) := c(E(H))$.

The distance function $c$ of a Euclidean TSP instance $G=(V,E)$ satisfies the triangle inequality. Therefore we have 
for any set of three vertices $x, y, z\in V(G)$:
\begin{equation}
c(x,y) ~+~ c(y, z) ~ ~\ge~ ~ c(x, z).
\end{equation}

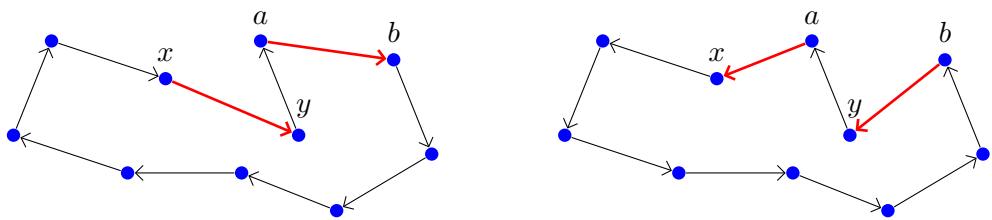
\begin{figure}[t]
\centering
\begin{tikzpicture}[scale=0.25]
\tikzstyle{vertex}=[blue,circle,fill, minimum size=5, inner sep=0]
\tikzstyle{arrow}=[Straight Barb[length=1mm]]

\node[vertex, label=above:$$] (P1)  at (17, 0) {};
\node[vertex, label=above:$b$] (P2)  at (20, 8) {};
\node[vertex, label=above:$$] (P3)  at ( 6, 2) {};
\node[vertex, label=above:$~y$] (P4)  at (15, 4) {};
\node[vertex, label=above:$$] (P5)  at (22, 3) {};
\node[vertex, label=above:$x$] (P6)  at ( 8, 7) {};
\node[vertex, label=above:$$] (P7)  at ( 0, 4) {};
\node[vertex, label=above:$$] (P8)  at (12, 2) {};
\node[vertex, label=above:$$] (P9)  at ( 2, 9) {};
\node[vertex, label=above:$a$] (P10) at (13, 9) {};

\draw[-{Straight Barb[length=1mm]},  red, line width=1]   (P10) to (P2);
\draw[-{Straight Barb[length=1mm]},  red, line width=1]   (P6)  to (P4);
\draw[-{Straight Barb[length=1mm]},  line width=0.4] (P4)  to (P10);
\draw[-{Straight Barb[length=1mm]},  line width=0.4] (P2)  to (P5);
\draw[-{Straight Barb[length=1mm]},  line width=0.4] (P5)  to (P1);
\draw[-{Straight Barb[length=1mm]},  line width=0.4] (P1)  to (P8);
\draw[-{Straight Barb[length=1mm]},  line width=0.4] (P8)  to (P3);
\draw[-{Straight Barb[length=1mm]},  line width=0.4] (P3)  to (P7);
\draw[-{Straight Barb[length=1mm]},  line width=0.4] (P7)  to (P9);
\draw[-{Straight Barb[length=1mm]},  line width=0.4] (P9)  to (P6);

\begin{scope}[shift={(29,0)}]
\node[vertex, label=above:$$] (P1)  at (17, 0) {};
\node[vertex, label=above:$b$] (P2)  at (20, 8) {};
\node[vertex, label=above:$$] (P3)  at ( 6, 2) {};
\node[vertex, label=above:$~y$] (P4)  at (15, 4) {};
\node[vertex, label=above:$$] (P5)  at (22, 3) {};
\node[vertex, label=above:$x$] (P6)  at ( 8, 7) {};
\node[vertex, label=above:$$] (P7)  at ( 0, 4) {};
\node[vertex, label=above:$$] (P8)  at (12, 2) {};
\node[vertex, label=above:$$] (P9)  at ( 2, 9) {};
\node[vertex, label=above:$a$] (P10) at (13, 9) {};

\draw[-{Straight Barb[length=1mm]},  red, line width=1]   (P10) to (P6);
\draw[-{Straight Barb[length=1mm]},  red, line width=1]   (P2)  to (P4);
\draw[-{Straight Barb[length=1mm]},  line width=0.4] (P4)  to (P10);
\draw[-{Straight Barb[length=1mm]},  line width=0.4] (P5)  to (P2);
\draw[-{Straight Barb[length=1mm]},  line width=0.4] (P1)  to (P5);
\draw[-{Straight Barb[length=1mm]},  line width=0.4] (P8)  to (P1);
\draw[-{Straight Barb[length=1mm]},  line width=0.4] (P3)  to (P8);
\draw[-{Straight Barb[length=1mm]},  line width=0.4] (P7)  to (P3);
\draw[-{Straight Barb[length=1mm]},  line width=0.4] (P9)  to (P7);
\draw[-{Straight Barb[length=1mm]},  line width=0.4] (P6)  to (P9);

\end{scope}

\end{tikzpicture}
\caption{An oriented TSP tour (left) and the tour obtained after replacing the edges  $(a,b)$ and $(x,y)$ with the edges $(a,x)$ and $(b,y)$ (right).
The orientation of the tour segment between the vertices $b$ and $x$ has been reversed in the new tour.}
\label{fig:2-Opt}
\end{figure}

The \2opt repeatedly replaces two edges from the tour by two other edges such that the resulting tour is shorter.
Given a tour $T$ and two edges $\{a,b\}$ and $\{x,y\}$ in $T$, there are two possibilities to replace these two edges by two other edges.
Either we can choose the pair $\{a,x\}$ and $\{b,y\}$ or we can choose the pair $\{a,y\}$ and $\{b,x\}$. Exactly one of these two pairs
will result in a tour again. Without knowing the other edges of $T$, we cannot decide which of the two possibilities is the correct one.
Therefore, we will assume in the following that the tour $T$ is an \emph{oriented} cycle, i.e., the edges of $T$ have an orientation 
such that each vertex has exactly one incoming and one outgoing edge. Using this convention, there is only one possibility to
exchange a pair of edges such that the new edge set is a tour again: two directed edges $(a,b)$ and $(x,y)$ have to be replaced by
the edges $(a,x)$ and $(b,y)$. Note that to obtain an oriented cycle again, one has to reverse the direction of the segment between $b$ and $x$, 
see Figure~\ref{fig:2-Opt}.

A TSP tour $T$ is called \emph{2-optimal} if for any two edges $(a,b)$ and $(x,y)$ of $T$ we have
\begin{equation}
 c(a,x) + c(b,y) ~\ge~ c(a,b) + c(x,y) 
 \label{eq:2-optimality}
\end{equation}
We call inequality~(\ref{eq:2-optimality}) the \emph{2-optimality condition}.

If $(a,b)$ and $(x,y)$ are two edges in a tour $T$ that violate the 2-optimality condition, i.e., they satisfy the inequality 
$ c(a,x) + c(b,y) < c(a,b) + c(x,y)$, 
then we can replace the edges $(a,b)$ and $(x,y)$ in $T$ by  the edges $(a,x)$ and $(b,y)$ and get a strictly shorter tour.
We call this operation of replacing the edges $(a,b)$ and $(x,y)$ in $T$ by  the edges $(a,x)$ and $(b,y)$
an \emph{improving 2-move}. Thus, the \2opt can be formulated as follows: \medskip

\framebox{\parbox{10cm}{
\noindent
{\bfseries 2-Opt Heuristic} ($V\subseteq \mathbb{R}^2$)\\[2mm]
1~ start with an arbitrary tour $T$ for $V$\\
2~ \texttt{while} there exists an improving 2-move in $T$ \\
3~ ~~~~  perform an improving 2-move\\
4~ \texttt{output} $T$}}\medskip

We call a Euclidean TSP instance  $V\subset \mathbb{R}^2$ \emph{degenerate} if there exists a line in $\mathbb{R}^2$ that
contains all points of $V$. Otherwise we call the instance \emph{non-degenerate}. 

It is easily seen that in a degenerate Euclidean TSP instance a 2-optimal tour is also an optimal tour:

\begin{proposition}
In a degenerate Euclidean TSP instance a 2-optimal tour is an optimal tour.
\label{prop:degenerate-case}
\end{proposition}
\begin{proof}
Let $V\subseteq \mathbb{R}^2$ be a degenerate Euclidean TSP instance. 
Let $c:V\times V \to \mathbb{R}$ be the Euclidean distance between two points in $V$.
Then there exist two points $a, b\in V$ such that the straight line segment $S$ from $a$ to $b$ contains all points of $V$. 
The length of an optimal TSP tour for $V$ is $2\cdot c(a,b)$. Assume there exists a $2$-optimal TSP tour $T$ that is not optimal. Orient the tour $T$. Then there must exist a point 
in $S\setminus V$ that is contained in at least three edges of the tour $T$ and therefore there must exist a point in $S\setminus V$ 
that is contained in two edges $(v,w)$ and $(x,y)$ of $T$ that are oriented in the same direction. This contradicts the 2-optimality of $T$ as 
$c(v,x) + c(w,y) < c(v,w) + c(x,y)$.   
\end{proof}

Because of Proposition~\ref{prop:degenerate-case} we may assume in the following that we have a non-degenerate Euclidean TSP instance. 

Let $T$ be a tour in a Euclidean TSP instance. Each edge of $T$ corresponds to a closed line segment in $\mathbb{R}^2$. 
A tour in a Euclidean TSP instance is called \emph{simple} if no two edges of the tour intersect in a point that lies in the interior of at least one
of the two corresponding line segments. 
For 2-optimal tours in Euclidean TSP instances we have the following simple but very important result. 

\begin{lemma}(Flood~\cite{Flo1956}) In a non-degenerate Euclidean TSP instance a 2-optimal tour is simple.
\label{lemma:nocrossing}
\end{lemma}

\subsection{Crossing-Free Pairs of Tours}
\label{sec:uncrossing}

\begin{figure}[t]
\centering
\begin{tikzpicture}[scale=0.43, fill=gray]

\coordinate  (P1) at (11, 12);
\coordinate  (P2) at ( 9,  3);
\coordinate  (P3) at ( 0,  7);
\coordinate  (P4) at (11,  4);
\coordinate  (P5) at ( 5, 13);
\coordinate  (P6) at ( 3, 15);
\coordinate  (P7) at ( 4, 12);
\coordinate  (P8) at (11,  2);
\coordinate  (P9) at ( 6,  9);
\coordinate (P10) at (10,  5);
\coordinate (P11) at (12, 14);
\coordinate (P12) at (14, 13);

\filldraw[fill=black!10!white, draw=red, line width = 1.5] 
(P1) -- (P11) -- (P12) -- (P10) -- (P4) -- (P8) -- (P2) -- (P9) -- (P3) -- (P7) -- (P6) -- (P5) -- cycle;

\draw[color=green!80!black, dash pattern = on 1.5mm off 1.5mm, line width = 1.5] 
(P1) -- (P4) -- (P8) -- (P2) -- (P10) -- (P3) -- (P7) -- (P6) -- (P5) -- (P9) -- (P11) -- (P12) -- cycle;

\tikzstyle{sal}=[-{Straight Barb[length=1.5mm]}, color=blue!70!white, line width = 1.5] 

\foreach \i in {1,...,12}
  \fill[black] (P\i) circle (2.5mm);

\begin{scope}[shift={(15.8,0)}]
\coordinate  (P1) at (11, 12);
\coordinate  (P2) at ( 9,  3);
\coordinate  (P3) at ( 0,  7);
\coordinate  (P4) at (11,  4);
\coordinate  (P5) at ( 5, 13);
\coordinate  (P6) at ( 3, 15);
\coordinate  (P7) at ( 4, 12);
\coordinate  (P8) at (11,  2);
\coordinate  (P9) at ( 6,  9);
\coordinate (P10) at (10,  5);
\coordinate (P11) at (12, 14);
\coordinate (P12) at (14, 13);

\filldraw[fill=black!10!white, draw=red, line width = 1.5] 
(P1) -- (P11) -- (P12) -- (P10) -- (P4) -- (P8) -- (P2) -- (P9) -- (P3) -- (P7) -- (P6) -- (P5) -- cycle;

\draw[color=green!80!black, dash pattern = on 1.5mm off 1.5mm, line width = 1.5] 
(P1) -- (P4) -- (P8) -- (P2) -- (P10) -- (P3) -- (P7) -- (P6) -- (P5) -- (P9) -- (P11) -- (P12) -- cycle;

\coordinate (P13) at (11.000000,  7.000000);
\coordinate (P14) at ( 9.833333, 12.194444); 
\coordinate (P15) at ( 7.777777,  5.444444);

\foreach \i in {1,...,12}
  \fill[black] (P\i) circle (2.5mm);

\foreach \i in {13,...,15}
  \fill[color=blue!70!white] (P\i) circle (2.5mm);

\end{scope}

\end{tikzpicture}
\caption{A Euclidean TSP instance with an optimal tour (red edges) and a 2-optimal tour (dashed green edges). Both tours shown in the left picture 
are simple but there are three pairs of crossing edges. The tours can be made crossing-free by adding three vertices (blue points in the right picture) to the instance.}
\label{fig:crossingEdges}
\end{figure}
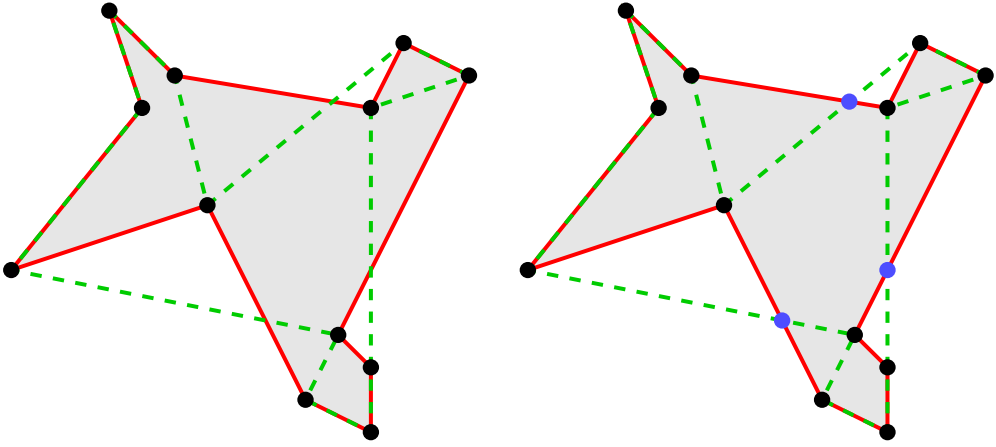

Let $T$ be an optimal tour and $T'$ be a 2-optimal tour in a non-degenerate Euclidean TSP instance. 
By Lemma~\ref{lemma:nocrossing} we know that both tours are simple. In the following we want to justify a much stronger assumption.
Two edges $e\in E(T)$ and $f\in E(T')$ \emph{cross} if $e$ and $f$ intersect in exactly one point in $\mathbb{R}^2$ 
and this point is in the interior of both line segments.  
We say that two tours $T$ and $T'$ are \emph{crossing-free} if there does not exist a pair of crossing edges.
See Figure~\ref{fig:crossingEdges} for an example of an optimal tour and a 2-optimal tour that have three crossing pairs of edges.

To prove Theorem~\ref{thm:main} it will be enough to prove it for the special case of 
crossing-free tours: 

\begin{theorem} Let $V\subseteq \mathbb{R}^2$ with $|V| = n$ be a non-degenerate Euclidean TSP instance, 
$T$ an optimal tour for $V$ and $S$ a 2-optimal tour for $V$.
If $T$ and $S$ are crossing-free then the length of $S$ is bounded by $O(\log n / \log \log n)$ times the length of $T$.
\label{thm:main-crossingfree}
\end{theorem}

The proof of Theorem~\ref{thm:main-crossingfree} will be presented in Section~\ref{sec:proof}. Here we show how Theorem~\ref{thm:main-crossingfree}
allows to prove Theorem~\ref{thm:main}. For this we describe a method to transform a pair of tours into a crossing-free pair of tours.

Let $V\subseteq \mathbb{R}^2$ be a Euclidean TSP instance and $T$ be a tour for $V$. We say that $V'\subseteq \mathbb{R}^2$ 
is a \emph{subdivision} for $(V,T)$ if $V \subset V'$ and $V'$ is a subset of the polygon $T$.  The set $V'$ \emph{induces} a new tour $T'$ which results from 
the tour $T$ by subdividing the edges by points in $V'\setminus V$. Note that $T$ and $T'$ constitute the same polygon.
Therefore we have:

\begin{proposition}
Let $V\subseteq \mathbb{R}^2$ be a Euclidean TSP instance and $T$ be an optimal tour. 
If $V'$ is a subdivision  for $(V,T)$ then the tour $T'$ induced by $V'$ is an optimal tour for $V'$.
\label{prop:optimality-preserved}
\end{proposition}

Subdividing a tour not only preserves the optimality but it also preserves the 2-optimality:

\begin{lemma}
Let $V\subseteq \mathbb{R}^2$ be a Euclidean TSP instance and $T$ be a 2-optimal tour. 
If $V'$ is a subdivision  for $(V,T)$ then the tour $T'$ induced by $V'$ is a 2-optimal tour for $V'$.
\label{lemma:2-optimality-preserved}
\end{lemma}

\begin{proof}
Let us assume that the tour $T'$ is oriented and that $(x',y')$ and $(a',b')$ are two edges of $T'$. We have to prove that 
these two edges satisfy the 2-optimality condition~(\ref{eq:2-optimality}). 
As $T'$ is a subdivision of the 2-optimal tour $T$ we know that there exist edges $(x,y)$ and $(a,b)$ in $T$ such that
the line segment $a'b'$ is contained in the line segment $ab$ and the line segment $x'y'$ is contained in the line segment 
$xy$. The 2-optimality of $T$ implies
\[ c(a,b) + c(x,y) ~\le~ c(a,x) + c(b,y) .\]
Using this inequality and the triangle inequality we get:
\begin{eqnarray*}
c(a',b') + c(x', y') & =   & c(a,b) - c(a,a') - c(b,b') + c(x,y) - c(x,x') - c(y,y') \\
                     & \le & c(a,x) - c(a,a') - c(x,x') + c(b,y) - c(b,b') - c(y,y') \\
                     & \le & c(a',x') + c(b',y') .
\end{eqnarray*}             
\end{proof}

Now we are able to reduce Theorem~\ref{thm:main} to Theorem~\ref{thm:main-crossingfree}:

\noindent
\textit{Proof of Theorem~\ref{thm:main}:~}
Let $V\subseteq \mathbb{R}^2$ be a Euclidean TSP instance with $|V| = n$,  $T$ be an optimal tour for $V$ and
$S$ be a 2-optimal tour for $V$. 
By Proposition~\ref{prop:degenerate-case} we may assume that $V$ is non-degenerate.
Let $V'\subseteq \mathbb{R}^2$ be the set of points obtained by adding to $V$ all
crossings between pairs of edges in $T$ and $S$. 
Denote the cardinality of $V'$ by $n'$. 
Let $T'$ and $S'$ be the tours induced by $V'$ for $T$ and $S$. 
Then by Proposition~\ref{prop:optimality-preserved}
and by Lemma~\ref{lemma:2-optimality-preserved} we know that $T'$ has the same length as $T$ and is an optimal tour for $V'$ 
and $S'$ has the same length as $S$ and is a 2-optimal tour for $V'$.
Now Theorem~\ref{thm:main-crossingfree} implies that the length of $S$ is at most 
$O(\log n' / \log \log n')$ times the length of $T$. It remains to observe that there can be at most $O(n^2)$ crossings between 
edges in $T$ and $S$ and therefore $O(\log n' / \log \log n') = O(\log n / \log \log n)$.
\qed
\medskip

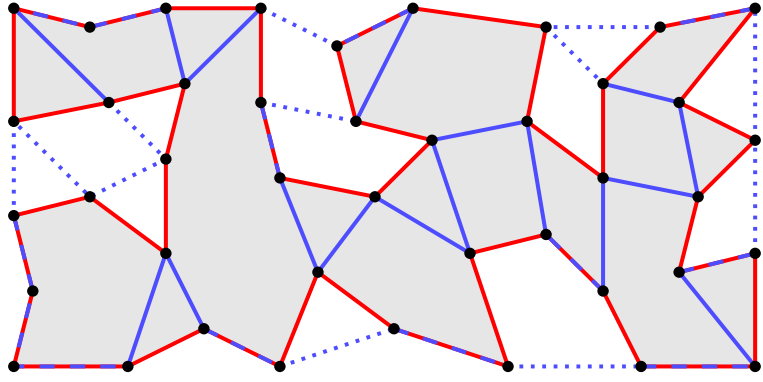
\begin{figure}[t]
\centering
\begin{tikzpicture}[scale=0.25, fill=gray]

\coordinate  (P1) at (32,  5);
\coordinate  (P2) at (18, 18);
\coordinate  (P3) at (27,  1);
\coordinate  (P4) at (15,  1);
\coordinate  (P5) at (23, 13);
\coordinate  (P6) at (40, 13);
\coordinate  (P7) at (34,  1);
\coordinate  (P8) at (15, 11);
\coordinate  (P9) at (11,  3);
\coordinate (P10) at (32, 11);
\coordinate (P11) at (35, 19);
\coordinate (P12) at ( 5, 10);
\coordinate (P13) at (21,  3);
\coordinate (P14) at (29, 19);
\coordinate (P15) at (25,  7);
\coordinate (P16) at (40,  7);
\coordinate (P17) at (36, 15);
\coordinate (P18) at (10, 16);
\coordinate (P19) at ( 1, 20);
\coordinate (P20) at (32, 16);
\coordinate (P21) at ( 9,  7);
\coordinate (P22) at (28, 14);
\coordinate (P23) at (36,  6);
\coordinate (P24) at ( 2,  5);
\coordinate (P25) at ( 7,  1);
\coordinate (P26) at (40,  1);
\coordinate (P27) at ( 1, 14);
\coordinate (P28) at ( 9, 20);
\coordinate (P29) at (20, 10);
\coordinate (P30) at (17,  6);
\coordinate (P31) at ( 6, 15);
\coordinate (P32) at (14, 15);
\coordinate (P33) at (22, 20);
\coordinate (P34) at (29,  8);
\coordinate (P35) at ( 1,  9);
\coordinate (P36) at (40, 20);
\coordinate (P37) at ( 9, 12);
\coordinate (P38) at (14, 20);
\coordinate (P39) at (37, 10);
\coordinate (P40) at ( 1,  1);
\coordinate (P41) at (19, 14);
\coordinate (P42) at ( 5, 19);

\filldraw[fill=black!10!white, draw=red, line width = 1.5] 
(P26) -- (P16) -- (P23) -- (P39) --  (P6) -- (P17) -- (P36) -- (P11) -- (P20) -- (P10) -- 
(P22) -- (P14) -- (P33) --  (P2) -- (P41) --  (P5) -- (P29) --  (P8) -- (P32) -- (P38) -- 
(P28) -- (P42) -- (P19) -- (P27) -- (P31) -- (P18) -- (P37) -- (P21) -- (P12) -- (P35) -- 
(P24) -- (P40) -- (P25) --  (P9) --  (P4) -- (P30) -- (P13) --  (P3) -- (P15) -- (P34) -- 
 (P1) --  (P7) -- cycle;

\draw[dash pattern = on 2mm off 2mm, color=blue!70!white, line width = 1.5] 
(P23) -- (P16)    (P36) -- (P11)     (P1) -- (P34)     (P8) -- (P32)    (P33) --  (P2) 
(P28) -- (P42) -- (P19)    (P35) -- (P24) -- (P40) -- (P25)     (P9) --  (P4)    (P13) -- 
 (P3)     (P7) -- (P26);

\draw[dash pattern = on 0.5mm off 1mm, color=blue!70!white, line width = 1.5] 
(P16) --  (P6) -- (P36)    (P11) -- (P14) -- (P20)    (P32) -- (P41)     (P2) -- (P38) 
(P31) -- (P37) -- (P12) -- (P27) -- (P35)     (P4) -- (P13)     (P3) --  (P7);

\draw[color=blue!70!white, line width = 1.5] 
(P26) -- (P23)    (P20) -- (P17) -- (P39) -- (P10) --  (P1)    (P34) -- (P22) --  (P5) -- 
(P15) -- (P29) -- (P30) --  (P8)    (P41) -- (P33)    (P38) -- (P18) -- (P28)    (P19) -- 
(P31)    (P25) -- (P21) --  (P9);

\foreach \i in {1,...,42}
  \fill[black] (P\i) circle (3mm);

\end{tikzpicture}
\caption{A Euclidean TSP instance with an optimal tour $T$ (red edges) and a 2-optimal tour 
(blue edges) that are crossing-free. The edges of the 2-optimal tour are partitioned into the 
edges lying in the interior of $T$ (solid blue lines), the edges that lie in the exterior of $T$
(dotted blue lines), and the edges that are part of $T$ (dashed blue edges).} 
\label{fig:crossingfreetours}
\end{figure}

\subsection{Partitioning the Edge Set of a 2-Optimal Tour}
\label{sec:edge-partition}

Let $V \subseteq\mathbb{R}^2$ be a non-degenerate Euclidean TSP instance, $T$ be an optimal tour and $S$ be a 2-optimal tour such that
$S$ and $T$ are crossing-free. As $S$ and $T$ are simple polygons and $S$ and $T$ are crossing-free we can partition the edge set of $S$
into three sets $S_1$, $S_2$, and $S_3$ such that all edges of $S_1$ lie in the interior of $T$, all edges of $S_2$ lie in the exterior of $T$
and all edges of $S_3$ are contained in $T$ (see Figure~\ref{fig:crossingfreetours}). 
More precisely, an edge $\{a, b\} \in S$ belongs to $S_1$ resp.\ $S_2$ if the corresponding open line segment $ab$ completely
lies in the interior resp.\ exterior of the polygon $T$. The set $S_3$ contains all the edges of $S$ that are subsets 
of the polygon $T$.

\begin{figure}[t]
\centering
\begin{tikzpicture}[scale=0.25, fill=gray]

\coordinate  (P1) at (32,  5);
\coordinate  (P2) at (18, 18);
\coordinate  (P3) at (27,  1);
\coordinate  (P4) at (15,  1);
\coordinate  (P5) at (23, 13);
\coordinate  (P6) at (40, 13);
\coordinate  (P7) at (34,  1);
\coordinate  (P8) at (15, 11);
\coordinate  (P9) at (11,  3);
\coordinate (P10) at (32, 11);
\coordinate (P11) at (35, 19);
\coordinate (P12) at ( 5, 10);
\coordinate (P13) at (21,  3);
\coordinate (P14) at (29, 19);
\coordinate (P15) at (25,  7);
\coordinate (P16) at (40,  7);
\coordinate (P17) at (36, 15);
\coordinate (P18) at (10, 16);
\coordinate (P19) at ( 1, 20);
\coordinate (P20) at (32, 16);
\coordinate (P21) at ( 9,  7);
\coordinate (P22) at (28, 14);
\coordinate (P23) at (36,  6);
\coordinate (P24) at ( 2,  5);
\coordinate (P25) at ( 7,  1);
\coordinate (P26) at (40,  1);
\coordinate (P27) at ( 1, 14);
\coordinate (P28) at ( 9, 20);
\coordinate (P29) at (20, 10);
\coordinate (P30) at (17,  6);
\coordinate (P31) at ( 6, 15);
\coordinate (P32) at (14, 15);
\coordinate (P33) at (22, 20);
\coordinate (P34) at (29,  8);
\coordinate (P35) at ( 1,  9);
\coordinate (P36) at (40, 20);
\coordinate (P37) at ( 9, 12);
\coordinate (P38) at (14, 20);
\coordinate (P39) at (37, 10);
\coordinate (P40) at ( 1,  1);
\coordinate (P41) at (19, 14);
\coordinate (P42) at ( 5, 19);

\filldraw[fill=black!10!white, draw=red, line width = 1.5] 
(P26) -- (P16) -- (P23) -- (P39) --  (P6) -- (P17) -- (P36) -- (P11) -- (P20) -- (P10) -- 
(P22) -- (P14) -- (P33) --  (P2) -- (P41) --  (P5) -- (P29) --  (P8) -- (P32) -- (P38) -- 
(P28) -- (P42) -- (P19) -- (P27) -- (P31) -- (P18) -- (P37) -- (P21) -- (P12) -- (P35) -- 
(P24) -- (P40) -- (P25) --  (P9) --  (P4) -- (P30) -- (P13) --  (P3) -- (P15) -- (P34) -- 
 (P1) --  (P7) -- cycle;

\foreach \i in {1,...,42}
  \node[circle, minimum size = 2.5mm, inner sep = 0mm] (N\i) at (P\i) {};

\node[circle, minimum size = 2.5mm, inner sep = 0mm, label=right:\hspace*{-1mm}$x_0$] (N26) at (P26) {};
\node[circle, minimum size = 2.5mm, inner sep = 0mm, label=left:$y_0$\hspace*{-1.6mm}]  (N23) at (P23) {};


\tikzstyle{sal}=[-{Straight Barb[length=1.5mm]}, color=blue!70!white, line width = 1.5] 

\tikzstyle{dal}=[dash pattern = on 0.5mm off 1mm, -{Straight Barb[length=1.5mm]}, color=blue!70!white, line width = 1.5] 

\tikzstyle{hl}=[dash pattern = on 0.1mm off 0.2mm, color=green!70!white, line width = 7] 

\draw[hl] (N26)  -- (N23); 
\draw[hl]  (N20) -- (N17); 
\draw[hl]  (N20) -- (N17); 
\draw[hl]  (N17) -- (N39); 
\draw[hl]  (N34) -- (N22); 
\draw[hl]  (N15) -- (N29); 
\draw[hl]  (N30) --  (N8); 
\draw[hl]  (N41) -- (N33); 
\draw[hl]  (N18) -- (N28); 
\draw[hl]  (N25) -- (N21); 

\draw[sal] (N26)  to node[left] {$e_0~$} (N23);
\draw[dal] (N23) -- (N16);
\draw[dal] (N16) -- (N6);
\draw[dal]  (N6) -- (N36);
\draw[dal] (N36) -- (N11);
\draw[dal] (N11) -- (N14);
\draw[dal] (N14) -- (N20);
\draw[sal] (N20) -- (N17);
\draw[sal] (N20) -- (N17);
\draw[sal] (N17) -- (N39);
\draw[sal] (N39) -- (N10);
\draw[sal] (N10) --  (N1);
\draw[dal]  (N1) -- (N34);
\draw[sal] (N34) -- (N22);
\draw[sal] (N22) --  (N5);
\draw[sal]  (N5) -- (N15);
\draw[sal] (N15) -- (N29);
\draw[sal] (N29) -- (N30);
\draw[sal] (N30) --  (N8);
\draw[dal]  (N8) -- (N32);
\draw[dal] (N32) -- (N41);
\draw[sal] (N41) -- (N33);
\draw[dal] (N33) --  (N2);
\draw[dal]  (N2) -- (N38);
\draw[sal] (N38) -- (N18);
\draw[sal] (N18) -- (N28);
\draw[dal] (N28) -- (N42);
\draw[dal] (N42) -- (N19);
\draw[sal] (N19) -- (N31);
\draw[dal] (N31) -- (N37);
\draw[dal] (N37) -- (N12);
\draw[dal] (N12) -- (N27);
\draw[dal] (N27) -- (N35);
\draw[dal] (N35) -- (N24);
\draw[dal] (N24) -- (N40);
\draw[dal] (N40) -- (N25);
\draw[sal] (N25) -- (N21);
\draw[sal] (N21) --  (N9);
\draw[dal]  (N9) --  (N4);
\draw[dal]  (N4) -- (N13);
\draw[dal] (N13) --  (N3);
\draw[dal]  (N3) --  (N7);
\draw[dal]  (N7) -- (N26);

\foreach \i in {1,...,42}
  \fill[black] (P\i) circle (3mm);

\end{tikzpicture}
\caption{The edge $e_0=(x_0,y_0)$ of the 2-optimal tour (blue edges) defines the set $S_1'$ of all green marked edges in the interior of the optimal tour $T$ (red edges).} 
\label{fig:edge-partition}
\end{figure}
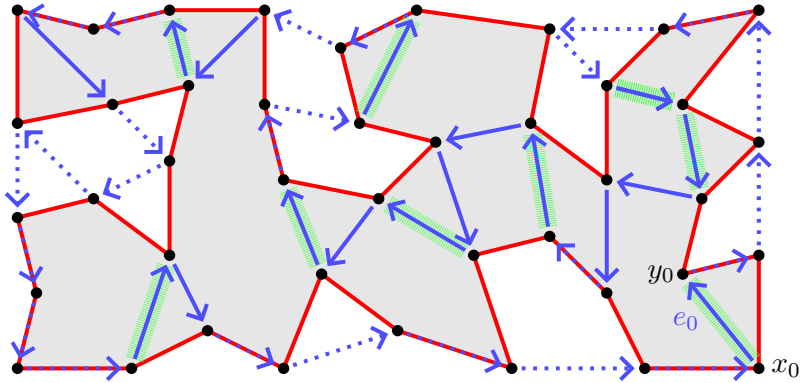

By definition we know that \blue{the total length} of all edges in $S_3$ \blue{is at most the length of the tour} $T$. 
To bound the total length of all edges in $S_1$ in terms of the length of $T$ we proceed as follows: Fix some 
orientation of
the tour $S$. We may assume that $S_1$ contains at least two edges as otherwise by the triangle inequality the length of $T$ is an upper bound for the
length of the edges in $S_1$. 
Choose an edge $e_0=(x_0,y_0)$ from $S_1$ such that one of the two $x_0$-$y_0$-paths in $T$ does not contain in its interior the endpoints of any
other edge in $S_1$ (see Figure~\ref{fig:edge-partition}).  

Let $T_{[x_0,y_0]}$ be the $x_0$-$y_0$-path in $T$ that contains the endpoints of all other edges in $S_1$. The path $T_{[x_0,y_0]}$ is unique if 
we assume $|S_1|\ge 2$. 
Then we define the set $S_1'$ to contain all edges from $S_1$ that are ``compatible'' with $T_{[x_0,y_0]}$ in the following sense:
\[ S_1' := \{(a,b) \in S_1 : \mbox{ the } x_0\mbox{-}b\mbox{-}\mbox{path in } T_{[x_0,y_0]} \mbox{ contains } a\} . \]
In Figure~\ref{fig:edge-partition} for the chosen edge $e_0=(x_0,y_0)$ the edges in $S_1'$ are marked green.

All edges in $S_1$ that are oriented in the ``wrong'' way with respect to $T_{[x_0,y_0]}$ define the set $S_1''$, i.e., we have
$S_1'' := S_1 \setminus S_1'$. Similarly we can define sets $S_2'$ and $S_2''$ with respect to some edge $f\in S$ that lies in the
exterior of $T$. We want to prove that for each of the four sets $S_1', S_1'', S_2', S_2''$ we can bound the total length of all edges by 
 $O(\log n / \log \log n)$ times the length of $T$. To achieve this we will reduce the problem to a problem in 
weighted arborescences.

\section{Arborescences and Pairs of Tours}
\label{sec:proofidea}

In this section we will explain why bounding the length of a 2-optimal tour
reduces to some problem in weighted arborescences. We start by giving an informal description of the idea. 

Let $T$ be an optimal tour for a non-degenerate Euclidean TSP instance $V\subseteq \mathbb{R}^2$ and let $S$ be a 2-optimal tour such that $S$ and $T$ are crossing-free. 
Then $T$ together with the edge set $S_1'$ as defined in Section~\ref{sec:edge-partition} is a plane graph. 
Each region of this plane graph is bounded by edges in $E(T) \cup S_1'$. The boundary of each region is a cycle. Because of the
triangle inequality we can bound the length of each edge in a cycle by the sum of the lengths of all other edges in this cycle. 
This way we get a bound for the length of each edge in $S_1'$ which we call the \emph{combined triangle inequality} as
it arises by applying the triangle inequality to edges from both tours $T$ and $S$.

From the boundaries of the regions of the plane graph we can derive
another type of inequalities. Suppose some boundary $B$ contains at least two edges from $S_1'$. Then there will be two distinct
edges $e,f\in B\cap S_1'$ such that $e$ and $f$ are oriented in opposite direction along the boundary $B$. 
(Here we see the reason why we partitioned the edge set $S_1$ into the two subsets $S_1'$ and $S_1''$. 
In the plane graph arising from $T$ together with the whole edge set $S_1$ it may happen, that the boundary of a region contains 
at least two edges from $S_1$ and these edges are all oriented in the same direction along the boundary.)
If we remove $e$ and $f$ from
$B$ we get two paths (one of which may be empty) connecting the heads and tails of $e$ and $f$. Now by applying 
the triangle inequality to these two paths and using the 2-optimality condition~(\ref{eq:2-optimality}) for the edges $e$ and $f$ 
we get another inequality for the edges in $S_1'$. We call this inequality the \emph{combined 2-optimality condition} as
it arises by applying the 2-optimality condition in combination with the triangle inequality to edges from both tours $T$ and $S$.
   
In the following we want to apply the combined triangle inequality and the combined 2-optimality condition to neighboring regions 
of the plane
graph formed by the edge set $T\cup S_1'$. This part of the proof is independent of the embedding induced by the 
point coordinates in $\mathbb{R}^2$. In particular this part of the proof can also be applied to point sets in higher 
dimension by choosing an arbitrary planar embedding of the tour $T$. We therefore reduce in the following the problem of bounding 
the total length of all edges in $S_1'$ to a purely combinatorial problem in weighted arborescences. \bigskip

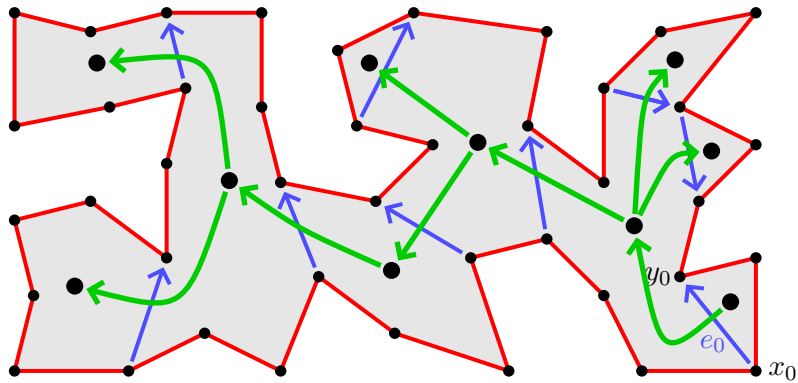
\begin{figure}[t]
\centering
\begin{tikzpicture}[scale=0.25, fill=gray]

\coordinate  (P1) at (32,  5);
\coordinate  (P2) at (18, 18);
\coordinate  (P3) at (27,  1);
\coordinate  (P4) at (15,  1);
\coordinate  (P5) at (23, 13);
\coordinate  (P6) at (40, 13);
\coordinate  (P7) at (34,  1);
\coordinate  (P8) at (15, 11);
\coordinate  (P9) at (11,  3);
\coordinate (P10) at (32, 11);
\coordinate (P11) at (35, 19);
\coordinate (P12) at ( 5, 10);
\coordinate (P13) at (21,  3);
\coordinate (P14) at (29, 19);
\coordinate (P15) at (25,  7);
\coordinate (P16) at (40,  7);
\coordinate (P17) at (36, 15);
\coordinate (P18) at (10, 16);
\coordinate (P19) at ( 1, 20);
\coordinate (P20) at (32, 16);
\coordinate (P21) at ( 9,  7);
\coordinate (P22) at (28, 14);
\coordinate (P23) at (36,  6);
\coordinate (P24) at ( 2,  5);
\coordinate (P25) at ( 7,  1);
\coordinate (P26) at (40,  1);
\coordinate (P27) at ( 1, 14);
\coordinate (P28) at ( 9, 20);
\coordinate (P29) at (20, 10);
\coordinate (P30) at (17,  6);
\coordinate (P31) at ( 6, 15);
\coordinate (P32) at (14, 15);
\coordinate (P33) at (22, 20);
\coordinate (P34) at (29,  8);
\coordinate (P35) at ( 1,  9);
\coordinate (P36) at (40, 20);
\coordinate (P37) at ( 9, 12);
\coordinate (P38) at (14, 20);
\coordinate (P39) at (37, 10);
\coordinate (P40) at ( 1,  1);
\coordinate (P41) at (19, 14);
\coordinate (P42) at ( 5, 19);

\filldraw[fill=black!10!white, draw=red, line width = 1.5] 
(P26) -- (P16) -- (P23) -- (P39) --  (P6) -- (P17) -- (P36) -- (P11) -- (P20) -- (P10) -- 
(P22) -- (P14) -- (P33) --  (P2) -- (P41) --  (P5) -- (P29) --  (P8) -- (P32) -- (P38) -- 
(P28) -- (P42) -- (P19) -- (P27) -- (P31) -- (P18) -- (P37) -- (P21) -- (P12) -- (P35) -- 
(P24) -- (P40) -- (P25) --  (P9) --  (P4) -- (P30) -- (P13) --  (P3) -- (P15) -- (P34) -- 
 (P1) --  (P7) -- cycle;

\foreach \i in {1,...,42}
  \node[circle, minimum size = 2.5mm, inner sep = 0mm] (N\i) at (P\i) {};

\node[circle, minimum size = 2.5mm, inner sep = 0mm, label=right:\hspace*{-1mm}$x_0$] (N26) at (P26) {};
\node[circle, minimum size = 2.5mm, inner sep = 0mm, label=left:$y_0$\hspace*{-1.6mm}]  (N23) at (P23) {};

\tikzstyle{arrow}=[Straight Barb[length=2mm]] 

\tikzstyle{sal}=[-{Straight Barb[length=1.5mm]}, color=blue!70!white, line width = 1.5] 

\draw[sal] (N26)  to node[below] {$e_0~$} (N23);
\draw[sal]  (N20) -- (N17); 
\draw[sal]  (N20) -- (N17);
\draw[sal]  (N17) -- (N39); 
\draw[sal]  (N34) -- (N22); 
\draw[sal]  (N15) -- (N29); 
\draw[sal]  (N30) --  (N8); 
\draw[sal]  (N41) -- (N33); 
\draw[sal]  (N18) -- (N28); 
\draw[sal]  (N25) -- (N21);

\coordinate  (D1) at ( 38.667,  4.667); 
\coordinate  (D2) at ( 33.600,  8.700); 
\coordinate  (D3) at ( 37.667, 12.667); 
\coordinate  (D4) at ( 35.750, 17.500); 
\coordinate  (D5) at ( 25.375, 13.125); 
\coordinate  (D6) at ( 19.667, 17.333); 
\coordinate  (D7) at ( 20.833,  6.333); 
\coordinate  (D8) at ( 12.300, 11.100); 
\coordinate  (D9) at (  5.333, 17.333); 
\coordinate (D10) at (  4.167,  5.500); 

\foreach \i in {1,...,10}
  \node[circle, minimum size = 3mm, inner sep = 0mm] (ND\i) at (D\i) {};
  
\tikzstyle{sadl}=[-{Straight Barb[length=1.5mm]}, color=green!80!black, line width = 2] 

\draw[sadl] (ND1) .. controls (35.0, 1.5) .. (ND2);
\draw[sadl] (ND2) .. controls (35.5, 12.5) .. (ND3);
\draw[sadl] (ND2) .. controls (34.0, 15.5) .. (ND4);
\draw[sadl] (ND2) -- (ND5);
\draw[sadl] (ND5) -- (ND6);
\draw[sadl] (ND5) -- (ND7);
\draw[sadl] (ND7) .. controls (16.0,  8.5) .. (ND8);
\draw[sadl] (ND8) .. controls (11.5, 18.0) .. (ND9);
\draw[sadl] (ND8) .. controls (10.0,  4.0) .. (ND10);

\foreach \i in {1,...,42}
  \fill[black] (P\i) circle (3mm);

\foreach \i in {1,...,10}
  \fill[color=black] (D\i) circle (4.5mm);

\end{tikzpicture}
\caption{The arborescence (green edges) in the dual of the plane graph formed by the edges of an optimal tour (red edges)
and the edges in the set $S_1'$ (blue edges) with respect to the edge $e_0=(x_0,y_0)$.} 
\label{fig:DualTree}
\end{figure}

We now give a formal description of the reduction. Consider the plane graph obtained from $T$ together with the edge set $S_1'$. 
Let $H$ be the graph that is obtained from
the geometric dual of this plane graph by removing the vertex corresponding to the outer region. Then each edge in $H$ is a dual 
of an edge in 
$S_1'$. As each edge in $S_1'$ is a chord in the polygon $T$ we know that each edge in $H$ is a cut edge and thus $H$ is a tree. 
See Figure~\ref{fig:DualTree} for an example.

We now want to orient the edges of $H$ to get an arborescence.
An \emph{arborescence} $A=(V,E)$ is a connected directed acyclic graph \blue{such that each vertex has at most one incoming edge. For 
a vertex $x\in V(A)$ we denote by $\delta^-(x)$ all incoming and by $\delta^+(x)$ all outgoing edges of $x$.}
Each arborescence has exactly one \emph{root} $r$ which is the unique vertex $r\in V(A)$ with $\delta^-(r) = \emptyset$. 
For an arborescence $A$ with root $r$ we say that $A$ \emph{is rooted at} $r$.

The set $S_1'$ has been defined with respect to some edge $e_0=(x_0,y_0)$. 
The tree $H$ contains a vertex that corresponds to the region of the plane graph that is bounded by the edge $e_0$ and 
the edges in $E(T) \setminus  T_{[x_0,y_0]}$. By choosing this vertex as the root and orienting all edges in $H$ from the root to the leaves,
we get an arborescence $A$ from the tree $H$ (see Figure~\ref{fig:DualTree}).

We want to define two weight functions on the edge set $E(A)$ of the arborescence $A$ to capture the weights of the edges in $T$ and the
edges in $S_1'$. 
First we define the function $c:E(A) \to \mathbb{R}_{>0}$ to be the weight of the corresponding dual edge in $S_1'$.
Secondly, we define  a weight function $w:E(A) \to \mathbb{R}_{>0}$ as follows. Let $e=(x,y)$ be a directed edge in $E(A)$.
Let $Y$ be the region corresponding to the vertex $y$ in the plane graph formed by $E(T) \cup S'_1$. 
Then we define $w(e)$ to be the weight of all the edges in $E(T)$ that belong to the boundary of $Y$.
\blue{Note that this definition implies $c(T) \ge w(A)$ and $c(S_1') = c(A)$.}

For the arborescence $A$ and the two weight functions $c$ and $w$ we can now state the above mentioned combined triangle inequality 
and the combined 2-optimality condition as follows:

\begin{lemma}\label{lemma:main-property-of-S1'}
Let $V\subseteq \mathbb{R}^2$ be a Euclidean TSP instance with distance function $\overline c: V\times V \to \mathbb{R}$ and
let $T$ be an optimal tour. Let $S$ be a 2-optimal tour such that $S$ and $T$ are crossing-free. 
Let $S_1'$ be defined as in Section~\ref{sec:edge-partition} with respect to some edge $e_0=(x_0,y_0)$. 
Let $A$ be the arborescence derived from the geometric dual of the plane graph $T\cup S_1'$ with 
weight functions $w:E(A) \to \mathbb{R}_{>0}$
and $c:E(A) \to \mathbb{R}_{>0}$ as defined above. Then 
we have
\begin{equation}
c(e) ~\le~ w(e) + \sum_{f\in \delta^+(y)} c(f) ~~~\mbox{for all $e= (x,y)\in E(A)$} .
\label{eq:combined-triangle-inequality}
\end{equation}
and 
\begin{equation}
c(x,y) + c(y,z) ~\le~ w(x,y) + \sum_{f\in \delta^+(y)\setminus\{(y,z)\}} c(f) \mbox{~~~~~ for all } (x,y), (y,z) \in E(A) 
\label{eq:combined-2-optimality}
\end{equation} 
\end{lemma}

\begin{proof}
Let $f=(a,b)$ be an edge in $S_1'$ and $f'=(a',b')$ its corresponding dual edge in $A$. By definition we have $c(f') = \overline c(f)$. 
The vertex $b'$ corresponds to a region $R$ in the plane graph on $V$ with edges $E(T) \cup S'_1$. 
By the triangle inequality the length $\overline c(f)$ of the edge $f$ is bounded by the length of all other edges in the boundary of the region $R$. 
Using the definitions of the functions $c$ and $w$ we therefore get: 
\[ c(f') ~=~ \overline c(f)  ~\le~ \sum_{g\in R\cap E(T)} \overline c(g) +  \sum_{g\in (R\cap S'_1) \setminus\{f\}} \overline c(g) 
         ~=~ w(f') + \sum_{g\in \delta^+(b')} c(g).\]
This proves~(\ref{eq:combined-triangle-inequality}).

We now prove property (\ref{eq:combined-2-optimality}). Let $f\in S'_1$. By definition of the set $S'_1$ we know that $S'_1$ is defined with 
respect to an edge $e_0=(x_0,y_0) \in S_1$ and the $x_0$-$y_0$-path $T_{[x_0,y_0]}$ contains the endpoints of all other edges in $S_1$. 
Let $\phi : V(T_{[x_0,y_0]}) \to\mathbb{N}$ such that $\phi(z)$ for $z \in  V(T_{[x_0,y_0]})$ denotes the distance 
(in terms of the number of edges) between $x_0$ and $z$ in $T_{[x_0,y_0]}$. The definition of the set $S'_1$ implies that $\phi(a) < \phi(b)$
for each edge $(a,b) \in S'_1$. Each edge in $S'_1$ can be seen as a shortcut for the path $T_{[x_0,y_0]}$. 
For an edge $f=(a,b)\in S'_1$ with dual edge $f'=(a',b') \in E(A)$ we denote by $\left(\delta^+(b')\right)'$
all edges dual to the edges in $\delta^+(b')$. The edge $f=(a,b)$ and the edges in $\left(\delta^+(b')\right)'$
belong to the border
of a region of the graph on $V$ with edge set $E(T) \cup S'_1$. Along this border the edge $f=(a,b)$ is directed opposite to all edges 
in $\left(\delta^+(b')\right)'$. Therefore, the triangle inequality together with the 2-optimality condition~(\ref{eq:2-optimality})
for the set $S'_1$ imply for each edge $(u,v)\in \left(\delta^+(b')\right)'$:
\[ \overline c(a,b) + \overline c (u,v) ~\le~ w(a',b') + \sum_{g\in \left(\delta^+(b')\right)' \setminus \{(u,v)\}} \overline c(g) \]
We have $\overline c(a,b) = c(a',b')$ and $\overline c(u,v) = c(b', x')$ for the vertex $x'\in V(A)$ such that $(u,v)\in S_1$ 
is the dual edge to $(b', x')\in E(A)$. 
Therefore we get:
\[ c(a',b') + c (b',x') ~\le~ w(a',b') + \sum_{g\in \delta^+(b') \setminus \{(b',x')\}} c(g) \]
\end{proof}

We call condition~(\ref{eq:combined-triangle-inequality}) the \emph{combined triangle inequality} and 
condition~(\ref{eq:combined-2-optimality}) the \emph{combined 2-optimality condition}.
Note that these two conditions can be formulated for any arborescence $A$ with weight functions $c$ and $w$. 
In the next section we will show that if these two conditions are satisfied for an arborescence $A$ then we can bound  $c(E(A))/w(E(A))$
by $O(\log(|E(A)|) / \log \log(|E(A)|))$.

\section{The Arborescence Lemmas}
\label{sec:arborescence-lemmas}

Let $A$ be an arborescence with weight functions $w:E(A) \to \mathbb{R}_{>0}$
and $c:E(A) \to \mathbb{R}_{>0}$. To simplify notation we set $w(A) := w(E(A))$ and $c(A) := c(E(A))$.
The main goal of this section is to prove a bound on the ratio $c(A)/w(A)$ if  
the combined triangle inequality~(\ref{eq:combined-triangle-inequality}) and the 
combined 2-optimality condition~(\ref{eq:combined-2-optimality}) hold.
To achieve this we will partition the edge set $E(A)$ into several subsets and bound the $c$-weight of these subsets 
in terms of $w(A)$. For a fixed number $l\in\mathbb{R}_{>0}$ we define a subset $E'\subseteq E(A)$ that contains all edges in $E(A)$ 
that have in some sense a small $c$-weight:

\begin{equation}
E' ~:=~ \{ (x,y) \in E(A):~~ c(x,y) ~<~ l \cdot \max_{f\in \delta^+(y)} c(f) \}.
\label{def:E'}
\end{equation}

We will later see how to choose the constant $l$. In the definition of $E'$ and in some of the following statements 
the maximum over of a possibly empty set appears. As usual we assume $\max \emptyset = -\infty$.

For the subset $E'\subseteq E(A)$ we can bound $c(E')$ in terms of $w(A)$ as follows:

\begin{lemma}
Let $A=(V,E)$ be an arborescence with weight functions $w:E(A)\to \mathbb{R}_{> 0}$ and $c:E(A)\to \mathbb{R}_{> 0}$
that satisfies the combined triangle inequality~(\ref{eq:combined-triangle-inequality}) and 
the combined 2-optimality condition~(\ref{eq:combined-2-optimality}).
Then we have: \[ c(E') ~\le~ \frac{l}{2} \cdot w(A) .\]
\label{lemma:E'}
\end{lemma}
\begin{proof}
The combined 2-optimality condition~(\ref{eq:combined-2-optimality}) states
$$c(x,y) + c(y,z) ~\le~ w(x,y) + \sum_{g\in \delta^+(y)\setminus\{(y,z)\}} c(g) \mbox{~~~~~ for all } (x,y), (y,z) \in E(A) $$

By adding $c(y,z) - c(x,y)$ to both sides of this inequality we get:
$$ 2 \cdot c(y,z) ~\le~ w(x,y) - c(x,y) + \sum_{g\in \delta^+(y)} c(g) \mbox{~~~~~ for all } (x,y), (y,z) \in E(A) $$

As this inequality holds for all $(y,z) \in \delta^+(y)$ it particularly holds for an edge $f\in\delta^+(y)$ that has maximum
$c$-weight:
$$ 2 \cdot \max_{f\in\delta^+(y)} c(f) ~\le~ w(x,y) - c(x,y) + \sum_{g\in \delta^+(y)} c(g) \mbox{~~~~~ for all } (x,y) \in E(A) $$
By definition of $E'$ we have $\displaystyle \frac1l \cdot c(x,y) <  \max_{f\in \delta^+(y)} c(f)$ for all edges $(x,y)\in E'$. 
This implies:

$$ \frac2l \cdot c(x,y) ~<~ w(x,y) - c(x,y) + \sum_{g\in \delta^+(y)} c(g) \mbox{~~~~~ for all } (x,y) \in E'$$

Adding this inequality for all $(x,y)\in E'$ and using that by the 
combined triangle inequality~(\ref{eq:combined-triangle-inequality})
we have $0  ~\le~ w(e) - c(e) + \sum_{f\in \delta^+(y)} c(f)$ for all $e= (x,y)\in E(A)$
we get:
\begin{eqnarray*}
\frac{2}{l}\cdot \sum_{(x,y)\in E'} c(x,y) & < &  \sum_{(x,y)\in E'}\left( w(x,y) - c(x,y) + \sum_{g\in \delta^+(y)} c(g)\right) \\
                                           &\le&  \sum_{(x,y)\in E(A)}\left( w(x,y) - c(x,y) + \sum_{g\in \delta^+(y)} c(g)\right) \\
                                           & \le & w(A)\vphantom{\left(\sum_x\right)}
\end{eqnarray*} 
\end{proof}

\begin{figure}[t]
\centering
\begin{tikzpicture}[scale=0.8]
\tikzstyle{vertex}=[black,circle,minimum size=8,inner sep=0]
\tikzstyle{arrow}=[-{Straight Barb[length=1.0mm]}]

\node[vertex, label = above:$r$]      (A)  at ( 3  , 4) {};
\node[vertex, red] (B)  at ( 2  , 3) {};
\node[vertex]      (C)  at ( 4  , 3) {};
\node[vertex, red] (D)  at ( 1  , 2) {};
\node[vertex]      (E)  at ( 3  , 2) {};
\node[vertex]      (F)  at ( 5  , 2) {};
\node[vertex, red] (G)  at ( 0  , 1) {};
\node[vertex, red] (H)  at ( 2  , 1) {};
\node[vertex, red] (I)  at ( 3  , 0) {};

\fill (A) circle (1mm);
\fill[red] (B) circle (1mm);
\fill (C) circle (1mm);
\fill[red] (D) circle (1mm);
\fill (E) circle (1mm);
\fill (F) circle (1mm);
\fill[red] (G) circle (1mm);
\fill[red] (H) circle (1mm);
\fill[red] (I) circle (1mm);

 \draw[arrow,  line width=1] (A) to (B);
 \draw[arrow,  line width=1] (A) to (C);
 \draw[arrow,  line width=1] (B) to (E);
 \draw[arrow,  line width=1] (C) to (F);
 
 \draw[arrow, red, line width=1] (D) to (G);
 \draw[arrow, red, line width=1] (D) to (H);
 \draw[arrow, red, line width=1] (H) to (I);
 \draw[arrow, red, line width=1] (B) to node[above left] {$e$} (D);
\end{tikzpicture}
\caption{An arborescence $A$ with root $r$. Shown in red is the sub-arborescence $A_e$ defined by the edge $e$.}
\label{fig:sub-arborescence}
\end{figure}
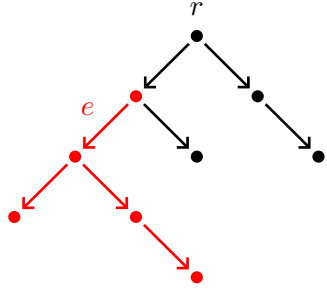

For an arborescence $A=(V,E)$ and an edge $e = (x,y)\in E(A)$ we denote by
$A_e$ the sub-arborescence rooted at $x$ that contains the edge $e$ and all descendants of $y$, see Figure~\ref{fig:sub-arborescence} for an example. The following lemma is a simple way to bound the $c$-weight of a single edge.

\begin{lemma}\label{lemma:weight-bound}
Let $A$ be an arborescence with weight functions $w:E(A) \to \mathbb{R}_{>0}$
and $c:E(A) \to \mathbb{R}_{>0}$ that satisfies  
the combined triangle inequality~(\ref{eq:combined-triangle-inequality}). 
Then we have 
\begin{equation}
c(e) ~\le~ w(A_e) ~~~\mbox{for all $e\in E(A)$}
\label{eq:weight-bound}
\end{equation}
\end{lemma}

\begin{proof}
This follows by induction on the height of the sub-arborescence $A_e$.
If $e=(x,y)$ is an edge in $E(A)$ such that $y$ is a leaf in $A$ then the combined triangle inequality~(\ref{eq:combined-triangle-inequality})
implies $c(e) \le w(e) = w(A_e)$. 
For an arbitrary edge $e=(x,y)\in E(A)$ we get 
from the combined triangle inequality~(\ref{eq:combined-triangle-inequality})
by induction: 
$$ c(e) ~\le~ w(e) + \sum_{f\in \delta^+(y)} c(f) ~\le~ w(e) + \sum_{f\in \delta^+(y)} w(A_f) ~=~ w(A_e) .$$
\end{proof}

Next we want to prove a statement similar to Lemma~\ref{lemma:E'} for other subsets of $E(A)$. 
For the fixed number $l$ that we chose in~(\ref{def:E'}) to define $E'$ and a number $r\in \mathbb{R}_{> 0}$ 
we define the edge set $E_r \subseteq E(A)$ as follows:
\begin{equation}
E_r ~:=~ \left\{ e=(x,y)\in E(A)\setminus E':~~ r ~<~ c(e) ~\le~ \frac{l}{4} \cdot r \right\}
\label{def:Er}
\end{equation}

\begin{lemma}\label{lemma:arborescence}
Let $A=(V,E)$ be an arborescence with weight functions $w:E(A)\to \mathbb{R}_{> 0}$ and $c:E(A)\to \mathbb{R}_{> 0}$
that satisfies the combined triangle inequality~(\ref{eq:combined-triangle-inequality})
and the combined 2-optimality condition~(\ref{eq:combined-2-optimality}).
Let $E_r$ be defined as in~(\ref{def:Er}).
Then we have: \[c(E_r) ~\le~ 2\cdot w(A) .\]  
\end{lemma}

\begin{proof}
Let $e = (x,y) \in E_r$. We first prove by induction on the cardinality of $E(A_e) \cap E_r$: 
\begin{equation}
w(A_e) ~\ge ~ c(e) + \sum_{f\in \left(E(A_e) \cap E_r\right) \setminus \{e\}} \left ( c(f) - \frac{r}{4}\right)
\label{eq:induction}
\end{equation}
 
If $|E(A_e) \cap E_r| = 1$ then $E(A_e) \cap E_r = \{e\}$ and therefore $(E(A_e) \cap E_r) \setminus \{e\} = \emptyset$.
Inequality~(\ref{eq:induction}) then states $w(A_e) \ge  c(e)$ which holds because of Lemma~\ref{lemma:weight-bound}.

Now assume that $|E(A_e) \cap E_r| > 1$ and that inequality~(\ref{eq:induction}) holds for all edges $f \in E_r$ with $|E(A_f) \cap E_r| < |E(A_e) \cap E_r|$. 
By the definition of the set $E_r$ we have $e=(x,y) \not\in E'$ and therefore by~(\ref{def:E'}) we know 
$c(x,y) \ge l \cdot c(f)$ for all $f\in\delta^+(y)$. Thus we get from~(\ref{def:E'}) and~(\ref{def:Er}) for each edge $f\in \delta^+(y)$: 
\begin{equation}
c(f) ~\le ~ \frac{1}{l}\cdot c(e)  ~\le~ \frac{1}{l}\cdot \frac{l}{4} \cdot r ~=~  \frac{r}{4}
\label{eq:r/4-bound}
\end{equation}
We define the following two sets of edges: 
\begin{equation*}
X ~ := ~ \{f\in \delta^+(y): E(A_f) \cap E_r = \emptyset\}
\end{equation*}
and
\begin{equation*}
F ~ := ~ \{f\in (E_r\cap E(A_e)) \setminus \{e\}: \mbox{ no edge } h\in E_r \mbox{ lies on a path from $f$ to $e$ in $A$} \}
\end{equation*}
For each edge $f\in \delta^+(y)$ we either have $E(A_f) \cap E_r = \emptyset$ or $E(A_f) \cap E_r \not= \emptyset$. In the first case the edge $f$ belongs to 
the set $X$.
In the second case at least one edge from $A_f$ belongs to $F$. Thus we have
\begin{equation}
|F| + |X| ~\ge~ |\delta^+(y)| ~~~\Rightarrow ~~~ |\delta^+(y) \setminus X| ~\le~ |F|. 
\label{eq:setsizes}
\end{equation}
 
By the induction hypothesis, inequality~(\ref{eq:induction}) holds for each edge $f\in F$ and we can now prove inequality~(\ref{eq:induction}) for the edge~$e$:
\begin{eqnarray*}
w(A_e) & \ge                                            & \sum_{f\in F} w(A_f) + \sum_{f\in X} w(A_f) + w(e) \\
       & \stackrel{(\text{\ref{eq:weight-bound}})}{\ge}  & \sum_{f\in F} w(A_f) + \sum_{f\in X} c(f) + w(e) \\
       & \stackrel{(\text{\ref{eq:combined-triangle-inequality}})}{\ge} & \sum_{f\in F} w(A_f) + c(e) - \sum_{f\in \delta^+(y)\setminus X} c(f) \\
       & \stackrel{(\text{\ref{eq:r/4-bound}})}{\ge}    & \sum_{f\in F} w(A_f) + c(e) - \sum_{f\in \delta^+(y)\setminus X} \frac{r}{4} \\
       & \stackrel{(\text{\ref{eq:setsizes}})}{\ge}     & \sum_{f\in F} \left( w(A_f) - \frac{r}{4} \right)  + c(e)  \\
       & \stackrel{(\text{\ref{eq:induction}})}{\ge}    & \sum_{f\in F} \left( c(f) + \sum_{g\in (E(A_f)\cap E_r)\setminus\{f\}} \left( c(g) -\frac{r}{4} \right) - \frac{r}{4} \right)  + c(e)  \\
       &                           =                    & c(e) + \sum_{f\in (E(A_e) \cap E_r) \setminus \{e\}} \left ( c(f) - \frac{r}{4}\right)  
\end{eqnarray*} 
The last equality holds because each edge in  $(E(A_e) \cap E_r) \setminus \{e\}$ appears exactly once in the sets 
$(E(A_f)\cap E_r)$ for $f\in F$.

By definition of the set $E_r$ we have for each edge $f\in E_r$: 
\begin{equation}
c(f) ~\ge~ r ~~~\Rightarrow~~~ \frac{1}{2} \cdot c(f) ~\ge~ \frac{r}{4} ~~~\Rightarrow~~~ c(f) - \frac{r}{4} ~\ge~ \frac{1}{2} \cdot c(f)
\label{eq:Er-edge-cost-lower-bound}
\end{equation} 
 
Inequality~(\ref{eq:induction}) therefore implies 
\begin{eqnarray*}
w(A_e) & \ge   & c(e) + \sum_{f\in (E(A_e) \cap E_r) \setminus \{e\}} \left ( c(f) - \frac{r}{4}\right)\\
       & \ge   & \sum_{f\in E(A_e) \cap E_r} \left ( c(f) - \frac{r}{4}\right) \\
       & \stackrel{(\text{\ref{eq:Er-edge-cost-lower-bound}})}{\ge}   & \sum_{f\in E(A_e) \cap  E_r} \left ( \frac{1}{2}  \cdot c(f)\right)\\
       & =     & \frac{1}{2} \cdot c(E(A_e) \cap E_r)  
\end{eqnarray*} 

Now choose a minimal set of edges $e_1, e_2, \ldots \in E_r$ such that $E_r \subseteq \bigcup_{i} E(A_{e_i})$.
Then 
\[ w(A) ~\ge~ w(\bigcup_{i} E(A_{e_i})) ~=~ \sum_{i} w(E(A_{e_i})) ~\ge~ \frac{1}{2} \cdot \sum_{i} c(E(A_{e_i}) \cap E_r)  ~=~ \frac{1}{2} \cdot c(E_r)\]
\end{proof}

Using the bounds in terms of $w(A)$ on $c(E')$  from Lemma~\ref{lemma:E'} and on $c(E_r)$ from Lemma~\ref{lemma:arborescence}
we can now prove the main result of this section:

\begin{lemma}
Let $A=(V,E)$ be an arborescence with weight functions $w:E(A)\to \mathbb{R}_{> 0}$ and $c:E(A)\to \mathbb{R}_{> 0}$
that satisfies the combined triangle inequality~(\ref{eq:combined-triangle-inequality})
and the combined 2-optimality condition~(\ref{eq:combined-2-optimality}).
Moreover we assume that $c(A) \ge 18 \cdot w(A)$.
Then we have: \[c(A) ~\le~ 12\cdot \frac{\log(|E(A)|)}{\log \log (|E(A)|)}\cdot w(A) .\]
\label{lemma:mainlemma}
\end{lemma}

\begin{proof}
We define $l := c(A) / w(A)$ and use this value in the definition~(\ref{def:E'}) of $E'$ and the definition~(\ref{def:Er})
of $E_r$. By assumption we have $l \ge 18$.
For $i= 1, 2, \ldots, \lfloor l/6 \rfloor$ we define $r_i := \left( \frac{4}{l}\right)^i\cdot w(A)$ and for these numbers we 
define sets $E_{r_i}$ as in~(\ref{def:Er}). By Lemma~\ref{lemma:weight-bound} we have  
$c(e) \le w(A) = \frac{l}{4} \cdot \left(\frac{4}{l}\right)^1\cdot w(A) = \frac{l}{4}\cdot r_1$ \blue{for all edges $e\in E(A)$} and therefore we have:
\[\bigcup_{i=1}^{\lfloor l/6 \rfloor} E_{r_i} = \left\{ e=(x,y)\in E(A)\setminus E': \left(\frac{4}{l}\right)^{\lfloor l/6 \rfloor} \cdot w(A) < c(e)\right\}.\]
Define \[E^* := \{e\in E(A): c(e) \le \left(\frac{4}{l}\right)^{\lfloor l/6 \rfloor} \cdot w(A)\} .\]
Then we have \[E(A) = E' \cup E^* \cup \bigcup_{i=1}^{\lfloor l/6 \rfloor} E_{r_i}.\] 

Using Lemma~\ref{lemma:E'} and Lemma~\ref{lemma:arborescence} we get:
\begin{eqnarray*}
l\cdot w(A) ~ = ~ c(A) & \le & \sum_{i=1}^{\lfloor l/6 \rfloor} c(E_{r_i}) + c(E') + c(E^*) \\
                       & \le & {\lfloor l/6 \rfloor} \cdot 2\cdot w(A) + \frac{l}{2} \cdot w(A) + \left(\frac{4}{l}\right)^{\lfloor l/6 \rfloor} \cdot w(A) \cdot|E^*| \\
                       & \le &  \frac{5}{6} \cdot l \cdot w(A) + \left(\frac{4}{l}\right)^{\lfloor l/6 \rfloor} \cdot w(A) \cdot|E^*|
\end{eqnarray*}

This implies 
\begin{equation}
|E(A)| ~\ge~ |E^*| ~\ge~ \frac{l}{6} \cdot \left(\frac{l}{4}\right)^{\lfloor l/6 \rfloor} ~\ge~ \left(\frac{l}{6}\right)^{l/6} .
\label{eq:sizeE}
\end{equation}

The function $\frac{\log x}{\log \log x}$ is monotone increasing for $x > 18$. Therefore we get from inequality~(\ref{eq:sizeE}): 

\begin{eqnarray*}
2\cdot \frac{\log(|E(A)|)}{\log\log(|E(A)|)}\cdot w(A) 
& \ge & 2\cdot \frac{\log\left(\left(\frac{l}{6}\right)^{l/6}\right)}{\log\log\left(\left(\frac{l}{6}\right)^{l/6}\right)}\cdot w(A)\\
& = & 2\cdot \frac{\frac{l}{6} \cdot \log\left(\frac{l}{6}\right)}{\log\left(\frac{l}{6}\right) + \log\log\left(\frac{l}{6}\right)}\cdot w(A)\\ 
& \ge & \frac{l}{6} \cdot w(A)\\
& = & \frac{1}{6} \cdot c(A)
\end{eqnarray*}

\end{proof}

\section{Proof of the Upper Bound for the 2-Opt Heuristic}
\label{sec:proof}

In this section we will prove Theorem~\ref{thm:main-crossingfree}.
Lemma~\ref{lemma:main-property-of-S1'} in combination with Lemma~\ref{lemma:mainlemma} shows that we can bound the length of
all edges in $S_1'$ by $O(\log n/\log \log n)$ times the length of an optimal tour $T$.
The statement of Lemma~\ref{lemma:main-property-of-S1'} also holds for the set $S''_1$: We can define an arborescence almost the same way as we did 
for the set $S'_1$ by taking the dual of the graph on $V$ formed by the edges of $T$ and $S''_1$ without the vertex for the outer region. The
only minor difference is the choice of the root vertex. For $S'_1$ we have chosen as root the vertex that corresponds to the region $R$
bounded by the edge $e_0=(x_0,y_0)$ and the edges in $E(T)\setminus T_{[x_0,y_0]}$. For the arborescence for $S''_1$ we choose as a root the region 
that contains \blue{the two vertices $x_0$ and $y_0$}.  The proof of Lemma~\ref{lemma:main-property-of-S1'} then without any changes shows that the statement of
Lemma~\ref{lemma:main-property-of-S1'} also holds for the set $S''_1$. Similarly, by exchanging the role of the outer and the inner region of $T$,
Lemma~\ref{lemma:main-property-of-S1'} also holds for the sets $S'_2$ and $S''_2$. We are now able to prove our main result:\medskip

\noindent
\textit{Proof of Theorem~\ref{thm:main-crossingfree}:~}
Let $V\subseteq \mathbb{R}^2$ with $|V| = n$ be a non-degenerate Euclidean TSP instance, 
$T$ an optimal tour for $V$ and $S$ a 2-optimal tour for $V$ such that $T$ and $S$ are crossing-free.
We partition the tour $S$ into the five (possibly empty) sets $S'_1$, $S''_1$, $S'_2$, $S''_2$, and $S_3$ as defined in Section~\ref{sec:edge-partition}.
Then $c(S_3) \le c(T)$. We claim that $c(S'_1) = O(\log n / \log \log n) \cdot c(T)$.
If $c(S'_1) < 18 \cdot c(T)$ this is certainly the case. Otherwise by Lemma~\ref{lemma:main-property-of-S1'} and 
Lemma~\ref{lemma:mainlemma} we get \blue{because of $c(T) \ge w(A)$ and $c(S_1') = c(A)$:}  
\[c(S'_1) ~\le~ 12\cdot \frac{\log(n)}{\log \log (n)}\cdot c(T) \]
which again proves the claim. As observed above, Lemma~\ref{lemma:main-property-of-S1'} also holds for the sets $S''_1$, $S'_2$, and $S''_2$. 
Therefore, we can apply the same argument to the sets $S''_1$, $S'_2$, and $S''_2$ and get
\[c(S) ~=~ c(S'_1) + c(S''_1) + c(S'_2) + c(S''_2) + c(S_3) ~=~  O(\log n / \log \log n) \cdot c(T).\]
\qed

\section{Proof of the Lower Bound for the \texorpdfstring{$k$}{k}-Opt heuristic} \label{sec:lower-bound}
\begin{figure}
\centering
\begin{tikzpicture}[scale=0.858]
\def\xscale{0.17}
\def\yscale{0.2}
\def\y#1{\yscale * \ifcase#1 0 \or 18 \or 24 \or 26 \fi}

\foreach \x in {0,3}
    {\draw[line width = 0.05mm] (0, \y\x) -- (81 * \xscale, \y\x);
     \draw (0,\y\x) node[left] {$l_\x$};}
     
\foreach \x in {1,2}
    {\draw[line width = 0.05mm] ( 0 * \xscale, \y\x) -- (27 * \xscale, \y\x);
     \draw[line width = 0.05mm] (54 * \xscale, \y\x) -- (81 * \xscale, \y\x);
     \draw (0,\y\x) node[left] {$l_\x$};}

\draw[line width = 0.05mm] (0, \y0) to node[left] {$h_0$} (0, \y1);
\draw[line width = 0.05mm] (81 * \xscale, \y0) to node[right] {$h'_0$} (81 * \xscale, \y1);

\draw[line width = 0.05mm] (27 * \xscale, \y1) to node[left]  {$h_1$}  (27 * \xscale, \y2);
\draw[line width = 0.05mm] (54 * \xscale, \y1) to node[right] {$h'_1$} (54 * \xscale, \y2);

\draw[line width = 0.05mm] ( 0 * \xscale, \y2) to node[left]  {}  ( 0 * \xscale, \y3);
\draw[line width = 0.05mm] (81 * \xscale, \y2) to node[right] {$h'_2$} (81 * \xscale, \y3);

\foreach \x in {0,..., 1}  \fill[color=green!80!black] (27 * \xscale * \x, \y0) circle(0.5mm);
\foreach \x in {0,..., 3}  \fill[color=green!80!black] ( 9 * \xscale * \x, \y1) circle(0.5mm);
\foreach \x in {0,..., 9}  \fill[color=green!80!black] ( 3 * \xscale * \x, \y2) circle(0.5mm);
\foreach \x in {0,...,27}  \fill[color=green!80!black] ( 1 * \xscale * \x, \y3) circle(0.5mm);

\foreach \x in {0,..., 1}  \fill[color=red] (54 * \xscale + 27 * \xscale * \x, \y0) circle(0.5mm);
\foreach \x in {0,..., 3}  \fill[color=red] (54 * \xscale +  9 * \xscale * \x, \y1) circle(0.5mm);
\foreach \x in {0,..., 9}  \fill[color=red] (54 * \xscale +  3 * \xscale * \x, \y2) circle(0.5mm);
\foreach \x in {0,...,27}  \fill[color=red] (54 * \xscale +  1 * \xscale * \x, \y3) circle(0.5mm);

\foreach \x in {1,...,26}  \fill (27 * \xscale +  1 * \xscale * \x, \y3) circle(0.5mm);

\foreach \y in {1,..., 17}  \fill[color=blue!70!white] ( 0 * \xscale, \y * \yscale) circle(0.5mm);
\foreach \y in {1,..., 17}  \fill[color=blue!70!white] (81 * \xscale, \y * \yscale) circle(0.5mm);
\foreach \y in {1,...,  5}  \fill[color=blue!70!white] (27 * \xscale, 18 * \yscale + \y * \yscale) circle(0.5mm);
\foreach \y in {1,...,  5}  \fill[color=blue!70!white] (54 * \xscale, 18 * \yscale + \y * \yscale) circle(0.5mm);
\fill[color=blue!70!white] ( 0 * \xscale, 24 * \yscale +  \yscale) circle(0.5mm);
\fill[color=blue!70!white] (81 * \xscale, 24 * \yscale +  \yscale) circle(0.5mm);

\end{tikzpicture}
  \caption{A structural drawing of the smallest non-trivial instance $I_3$ for $p=1$ and the tour $T$. 
For $p=1$ the instance $I_3$ already contains 2916 points. We therefore only draw a subset of the points and adapted the coordinates.  
The vertices in $V_1$ (green points), $V_2$ (red points), and $V_3$ (black points) lie on the horizontal lines $l_0,l_1,l_2$, and $l_3$ where $l_0$ is the bottommost line. The vertices in $V_4$ (blue points) lie on the vertical line segments $h_i$ and $h'_i$. }
  \label{euclidean tour}
\end{figure}
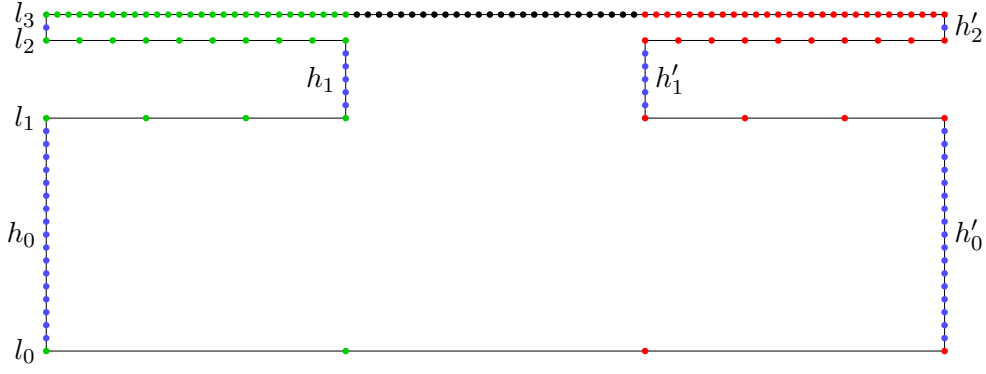

In this section we prove Theorem~\ref{thm:k-Opt}. For this we modify a construction of instances given in~\cite{CKT1999} which shows an asymptotic lower bound of $\frac{\log(n)}{\log\log(n)}$ for the 2-Opt heuristic. Our new instances yield the same asymptotic lower bound on the approximation ratio for the $k$-Opt heuristic as for the 2-Opt heuristic. As a generalization the lower bound not only works for the Euclidean TSP but for all TSP instances where the distances arise from the 
$p$-norm for some $p$ \blue{with $1\le p < \infty$}. Together with the matching upper bound from Theorem~\ref{thm:main} this implies that the $k$-Opt heuristic has an asymptotic approximation ratio of $\Theta\left(\frac{\log(n)}{\log\log(n)}\right)$ for Euclidean TSP. From now on, let us consider the $k$-Opt heuristic with the $p$-norm for some fixed $k,p$ with $k\geq 2$ and $p\geq 1$. We denote the $p$-norm by $\Vert \cdot\Vert_p$.
For every odd $q\in \mathbb{N}$ we construct an instance $I_q$ with 
\begin{align*}
n:=2\sum_{i=0}^q (q^{(p+1)i}+1)+q^{(p+1)q}-1+2\sum_{i=0}^{q-1} (q^{(p+1)(q-i)-1}-1)
\end{align*}
vertices (Figure~\ref{euclidean tour}). Note that $n\in\Theta\left(q^{(p+1)q}\right)$ and hence $q\in\Theta\left(\frac{\log n}{W(\log n)}\right)=\Theta\left(\frac{\log n}{\log \log n}\right)$ where $W$ is the Lambert $W$ function and we use the property of the Lambert $W$ function that the equation $x^x=z$ has the solution $x = \ln z /W(\ln z)$ and $W(\ln z) = \Theta(\ln \ln z)$. 

For the construction of $I_q$ first define $q+1$ lines $(l_i)_{i\in \{0,\dots, q\}}$ parallel to the $x$-axis. To define their $y$-coordinates we
use the definition
\begin{equation}
S(i) ~:=~ \sum_{s=0}^{i-1} q^{(p+1)(q-s)-1}
\end{equation}
The lines $l_i$ satisfy the function $y=S(i)$.  We will call $l_i$ the $i$th \emph{layer}.

The instance $I_q$ consists of four sets of vertices: $V_1,V_2,V_3$ and $V_4$.
For $V_1$ we place $\frac{q^{(p+1)q}}{q^{(p+1)(q-i)}}+1=q^{(p+1)i}+1$ equidistant vertices on the $i$th layer $l_i$ between the $x$-coordinate 0 and $q^{(p+1)q}$ such that the distance between consecutive vertices is $q^{(p+1)(q-i)}$. Note that the coordinates of these vertices are independent of the $p$-norm since the vertices lie on a line parallel to the $x$-axis.

The vertices in $V_2$ are copies of $V_1$ shifted to the right by $2q^{(p+1)q}$, i.e.\ every vertex in $V_1$ with coordinates $(e,f)$ corresponds to a vertex in $V_2$ with coordinates $(e + 2q^{(p+1)q},f)$. 

Now, we fill the gaps in the topmost layer $l_q$. The set $V_3$ consists of $q^{(p+1)q}-1$ vertices dividing the line segment between $(q^{(p+1)q}, S(q))$ and $(2q^{(p+1)q},S(q))$ into $q^{(p+1)q}$ equidistant parts such that the distance between consecutive vertices is~1.

Define the vertical line segments $(h_i)_{0\leq i <q}$ parallel to the $y$-axis with $y$-coordinate between  $S(i)$ and $S(i+1)$ as follows: If $i$ is even, it is the line segment with the $x$-coordinate 0, otherwise it is the line segment with the $x$-coordinate $q^{(p+1)q}$. 

Similarly, define the shifted vertical line segments $(h'_i)_{0\leq i <q}$ parallel to the $y$-axis with $y$-coordinate between $S(i)$ and $S(i+1)$ as follows: If $i$ is even, it is the line segment with the $x$-coordinate $3q^{(p+1)q}$, otherwise it is the line segment with the $x$-coordinate $2q^{(p+1)q}$.

Finally, the set $V_4$ consists of the following vertices: For each $i$ with $0\le i < q$ we place $q^{(p+1)(q-i)-1}-1$ equidistant vertices on $h_i$ and $h_i'$ such that the distance between two consecutive vertices is~1.

The coordinates of the vertices of the instance $I_q$ are given explicitly by:
\begin{align*}
V_1 :=&\bigcup_{0\leq i\leq q, 0\leq j \leq q^{(p+1)i}} \left\{(jq^{(p+1)(q-i)},S(i))\right\}\\
V_2:=&\bigcup_{0\leq i\leq q, 0\leq j \leq q^{(p+1)i}} \left\{(jq^{(p+1)(q-i)}+2q^{(p+1)q}, S(i))\right\}\\
V_3:=&\bigcup_{1\leq j\leq q^{(p+1)q}-1} \left\{(q^{(p+1)q}+j,S(q))\right\}\\
V_4:=&\bigcup_{\substack{0\leq i \leq q-1, i \text{ even},\\ 1\leq j\leq q^{(p+1)(q-i)-1}-1}} \left\{(0,j+S(i))\right\} \cup \left\{(3q^{(p+1)q},j+S(i))\right\}\\
&\bigcup_{\substack{0\leq i \leq q-1, i \text{ odd},\\ 1\leq j\leq q^{(p+1)(q-i)-1}-1}} \left\{(q^{(p+1)q},j+S(i))\right\} \cup \left\{(2q^{(p+1)q},j+S(i))\right\}
\end{align*}

Let $V(I_q):= V_1\cup V_2 \cup V_3 \cup V_4$. Note that $\lvert V_1 \rvert=\lvert V_2 \rvert= \sum_{i=0}^q (q^{(p+1)i}+1), \lvert V_3 \rvert=q^{(p+1)q}-1$ and $\lvert V_4 \rvert=2\sum_{i=0}^{q-1} (q^{(p+1)(q-i)-1}-1)$.
Hence,
\begin{align*}
\lvert V_1 \rvert+ \lvert V_2 \rvert + \lvert V_3 \rvert +\lvert V_4 \rvert&=2\sum_{i=0}^q (q^{(p+1)i}+1)+q^{(p+1)q}-1+2\sum_{i=0}^{q-1} (q^{(p+1)(q-i)-1}-1)\\
                                                                           &=n.
\end{align*}

Define the tour $T$ as shown in Figure~\ref{euclidean tour} by connecting consecutive equidistant vertices in $V_1,V_2,V_3$ and $V_4$.
More formally, define
\begin{align*}
E_1:=&\bigcup_{0\leq i\leq q, 0\leq j \leq q^{(p+1)i}-1} \left\{\{(jq^{(p+1)(q-i)},S(i)),((j+1)q^{(p+1)(q-i)},S(i))\}\right\}\\
E_2:=&\bigcup_{0\leq i\leq q, 0\leq j \leq q^{(p+1)i}-1} \left\{\{(jq^{(p+1)(q-i)}+2q^{(p+1)q}, S(i)),\right. \\
&\left.\hspace*{33mm}((j+1)q^{(p+1)(q-i)}+2q^{(p+1)q},S(i))\}\right\}\\
E_3:=&\bigcup_{0\leq j\leq q^{(p+1)q}-1} \left\{\{(q^{(p+1)q}+j,S(q)),(q^{(p+1)q}+j+1,S(q))\}\right\}\\
E_4:=&\bigcup_{\substack{0\leq i \leq q-1, i \text{ even},\\ 0\leq j\leq q^{(p+1)(q-i)-1}-1}} \left\{\{(0,j+S(i)),(0,j+1+S(i))\}\right\}\\
&\cup \left\{\{(3q^{(p+1)q},j+S(i)),(3q^{(p+1)q},j+1+S(i))\}\right\}\\
&\bigcup_{\substack{0\leq i \leq q-1, i \text{ odd},\\ 0\leq j\leq q^{(p+1)(q-i)-1}-1}} \left\{\{(q^{(p+1)q},j+S(i)),(q^{(p+1)q},j+1+S(i))\}\right\}\\
&\cup \left\{\{(2q^{(p+1)q},j+S(i)),(2q^{(p+1)q},j+1+S(i))\}\right\}\\
E_5:=&\left\{\{(q^{(p+1)q},0),(2q^{(p+1)q},0)\}\right\}
\end{align*}
Now, let $E(T):=E_1\cup E_2 \cup E_3\cup E_4 \cup E_5$.
Note that since $q+1$ is even, $T$ is indeed a tour. Let $T^*$ be an optimal tour of the instance $I$. Next, we bound the length of $T$ and $T^*$.

\begin{lemma} \label{k-Opt Euc T bound}
The length of the tour $T$ as defined above for the instance $I_q$ is at least $q\cdot q^{(p+1)q}$.
\end{lemma}

\begin{proof}
The proof is similar to the proof of Claim 4.5 in~\cite{CKT1999}.\\
Consider only the horizontal edges connecting consecutive vertices of $V_1$, i.e.\ the edge set $E_1$. On each of the $q+1$ layers these edges form line segments each with length $q^{(p+1)q}$. Hence, we can bound the length of the tour by $(q+1)\cdot q^{(p+1)q}>q\cdot q^{(p+1)q}$.
\end{proof}

\begin{lemma} \label{k-Opt Euc opt bound}
The length of the optimal tour $T^*$ for the instance $I_q$ is at most $14q^{(p+1)q}$.
\end{lemma}

\begin{proof}
The proof is similar to the proof of Claim 4.4 in~\cite{CKT1999}.\\
We bound the length of the optimal tour by twice the length of a spanning tree. For any vertex in $V_1 \cup V_2$ not on the topmost layer $l_q$ consider the vertical line segment to the next higher layer. Since the distance between consecutive vertices on $l_{i+1}$ is divisible by the distance of vertices on $l_i$, every vertex not on $l_p$ is connected this way with a vertex on the next higher layer. For all $i$ it is easy to see that $h_i$ and $h_i'$ and hence the vertices of $V_4$ lie on these vertical line segments. There are $2\left(q^{(p+1)i}+1\right)$ vertices on $l_i$ and each of the connection edges to $l_{i+1}$ has length $q^{(p+1)(q-i)-1}$. Thus, these edges have a total length of $2\sum_{i=0}^{q-1}q^{(p+1)(q-i)-1}\left(q^{(p+1)i}+1\right)$. We get a spanning tree by adding edges connecting consecutive vertices on $l_q$. These edges form a line segment with length $3q^{(p+1)q}$. Altogether, the total length of the spanning tree is:
\begin{align*}
&3q^{(p+1)q}+2\sum_{i=0}^{q-1}q^{(p+1)(q-i)-1}\left(q^{(p+1)i}+1\right)=3q^{(p+1)q}+2\sum_{i=0}^{q-1}\left(q^{(p+1)q-1}+q^{(p+1)(q-i)-1}\right)\\
\leq &3q^{(p+1)q}+ 4q^{(p+1)q}=7q^{(p+1)q}.
\end{align*}
The length of the optimal tour can now be bounded by twice the cost of this spanning tree.
\end{proof}

Combining both lemmas we can already see that the ratio between the length of $T$ and the optimal tour is at least $\frac{q\cdot q^{(p+1)q}}{14q^{(p+1)q}}=\frac{q}{14}$ and recall that $q\in \Theta\left(\frac{\log(n)}{\log\log(n)}\right)$. It remains to show that the tour $T$ for the instance $I_q$ is $k$-optimal for $q$ large enough. For that we first show some auxiliary lemmas.

The \emph{bounding box} of a set of points $P$ is the smallest axis-parallel rectangle containing all points in $P$. 

\begin{lemma}\label{connecting edge}
Let $P$ be a polygon such that the bounding box of the vertices of $P$ has the side length $d_x$ and $d_y$ in $x$ and $y$ direction, respectively. Then, the perimeter of $P$ is at least $2\sqrt[p]{d_x^p+d_y^p}$ where the distances are induced by the $p$-norm.
\end{lemma}

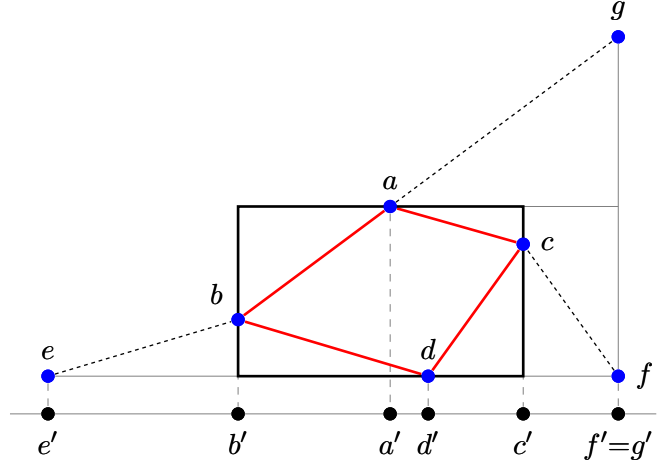
\begin{figure}[!htb]
\centering
\begin{tikzpicture}[scale=0.25]
\def\yb{3}
\def\xd{10}
\def\xa{8}
\def\yc{7}
\tikzstyle{vertex}=[circle,fill, minimum size=5, inner sep=0]
\tikzstyle{arrow}=[Straight Barb[length=1mm]]

\node[blue, vertex, label=above:$a$] (a)  at ( \xa,   9) {};
\node[blue, vertex, label=  135:$b$] (b)  at (   0, \yb) {};
\node[blue, vertex, label=right:$c$] (c)  at (  15, \yc) {};
\node[blue, vertex, label=above:$d$] (d)  at ( \xd,   0) {};
\node[blue, vertex, label=above:$e$] (e)  at (-\xd,   0) {};
\node[blue, vertex, label=right:$f$] (f)  at (15 + 15 - \xd,   0) {};
\node[blue, vertex, label=above:$g$] (g)  at (15 + 15 - \xd,   9 + 9) {};

\node[vertex, label=below:$a'$] (a')  at ( \xa,   -2) {};
\node[vertex, label=below:$b'$] (b')  at (   0,   -2) {};
\node[vertex, label=below:$c'$] (c')  at (  15,   -2) {};
\node[vertex, label=below:$d'$] (d')  at ( \xd,   -2) {};
\node[vertex, label=below:$e'$] (e')  at (-\xd,   -2) {};
\node[vertex, label=below:$f'{=}g'$] (f')  at (15 + 15 - \xd,   -2) {};

\draw[gray, line width = 0.25] (e) -- (f); 
\draw[gray, line width = 0.25] (g) -- (f); 
\draw[gray, line width = 0.25] (a) -- (15+15-\xd, 9); 
\draw[gray, line width = 0.25] (-\xd-2, -2) -- (15+15-\xd + 2, -2); 

\draw[gray, dash pattern = on 1mm off 1mm, line width = 0.25] (a) -- (a'); 
\draw[gray, dash pattern = on 1mm off 1mm, line width = 0.25] (b) -- (b'); 
\draw[gray, dash pattern = on 1mm off 1mm, line width = 0.25] (c) -- (c'); 
\draw[gray, dash pattern = on 1mm off 1mm, line width = 0.25] (d) -- (d'); 
\draw[gray, dash pattern = on 1mm off 1mm, line width = 0.25] (e) -- (e'); 
\draw[gray, dash pattern = on 1mm off 1mm, line width = 0.25] (f) -- (f');

\draw[red, line width=1]   (a) -- (b) -- (d) -- (c) -- (a);

\draw[line width = 1] (0,0) -- (15,0) -- (15,9) -- (0,9) -- cycle;

\draw[dash pattern = on 0.5mm off 0.5mm, line width = 0.5] (e) -- (b); 
\draw[dash pattern = on 0.5mm off 0.5mm, line width = 0.5] (f) -- (c); 
\draw[dash pattern = on 0.5mm off 0.5mm, line width = 0.5] (a) -- (15+15-\xd, 9 + 9);

\node[blue, vertex, label=above:$a$] (a)  at ( \xa,   9) {};
\node[blue, vertex, label=  135:$b$] (b)  at (   0, \yb) {};
\node[blue, vertex, label=right:$c$] (c)  at (  15, \yc) {};
\node[blue, vertex, label=above:$d$] (d)  at ( \xd,   0) {};
\node[blue, vertex, label=above:$e$] (e)  at (-\xd,   0) {};
\node[blue, vertex, label=right:$f$] (f)  at (15 + 15 - \xd,   0) {};
\node[blue, vertex, label=above:$g$] (g)  at (15 + 15 - \xd,   9 + 9) {};

\node[vertex, label=below:$a'$] (a')  at ( \xa,   -2) {};
\node[vertex, label=below:$b'$] (b')  at (   0,   -2) {};
\node[vertex, label=below:$c'$] (c')  at (  15,   -2) {};
\node[vertex, label=below:$d'$] (d')  at ( \xd,   -2) {};
\node[vertex, label=below:$e'$] (e')  at (-\xd,   -2) {};
\node[vertex, label=below:$f'{=}g'$] (f')  at (15 + 15 - \xd,   -2) {};
\end{tikzpicture}
  \caption{Sketch for the proof of Lemma~\ref{connecting edge} showing the short cutted polygon $P$ in red and the bounding box in black.}
  \label{sketch connecting edge}
\end{figure}

\begin{proof}
For each side of the bounding box mark a vertex of $P$ that lies on that side. Note that a vertex may be marked multiple times for different sides. For every other unmarked vertex we can shortcut the two adjacent edges to get a new polygon without increasing the perimeter and changing the bounding box. In the end we end up with a polygon that has at most 4 sides. We can further assume that the polygon is simple since otherwise we can perform a 2-move to remove the crossing edges without increasing the length of the perimeter and changing the bounding box. Therefore, we may assume that $P$ consists of the vertices $a,b,c,d$ lying on the top, left, right, and bottom side of the bounding box, respectively. Note that some of these vertices may coincide in case that $P$ contains less than four edges. We reflect the vertex $d$ by the left and right side of the bounding box to obtain $e$ and $f$, respectively. Then, we reflect $f$ by the top side of the bounding box to obtain $g$ (Figure~\ref{sketch connecting edge}). By symmetry and the triangle inequality, the perimeter of $P$ is at least.
\begin{align*}
&\Vert a-b\Vert_p+\Vert a-c\Vert_p+\Vert b-d\Vert_p+\Vert c-d\Vert_p\\
=&\Vert a-b\Vert_p+\Vert a-c\Vert_p+\Vert b-e\Vert_p+\Vert c-f\Vert_p\\
\geq &\Vert a-e\Vert_p+\Vert a-f\Vert_p=\Vert a-e\Vert_p+\Vert a-g\Vert_p\geq \Vert e-g\Vert_p
\end{align*}
Let $a',b',c',d',e',f'$ and $g'$ be the projections of $a,b,c,d,e,f$ and $g$ to the $x$-axis, respectively. Again by symmetry we have
\begin{align*}
\Vert e'-g'\Vert_p&=\Vert e'-a'\Vert_p+\Vert a'-g'\Vert_p=\Vert e'-b'\Vert_p+\Vert b'-a'\Vert_p+\Vert a'-f'\Vert_p\\
&=\Vert d'-b'\Vert_p+\Vert b'-a'\Vert_p+\Vert a'-c'\Vert_p+\Vert c'-f'\Vert_p\\
&=\Vert d'-b'\Vert_p+\Vert b'-a'\Vert_p+\Vert a'-c'\Vert_p+\Vert c'-d'\Vert_p=2d_x
\end{align*}
Together with a similar calculation with the projections of the vertices to the $y$-axis we can conclude that the bounding box of $\{e,g\}$ has side length $2d_x$ and $2d_y$. Therefore, we have $\Vert e-g\Vert_p=\sqrt[p]{(2d_x)^p+(2d_y)^p}=2\sqrt[p]{d_x^p+d_y^p}$ which completes the proof.
\end{proof}

\begin{lemma}\label{estimate}
For fixed $k$ and $p$ there is a $\bar{q}$ such that for all $q\geq \bar{q}$ and $0\leq a,b\leq k$ we have
\begin{align*}
\sqrt[p]{(aq^{(p+1)(q-s)})^p+q^{p((p+1)(q-s)-1)}}>a q^{(p+1)(q-s)}+bq^{(p+1)(q-s-1)}.
\end{align*}
\end{lemma}

\begin{proof}
We have for $q$ large enough
\begin{align*}
&\left(\sqrt[p]{(a q^{(p+1)(q-s)})^p+q^{p((p+1)(q-s)-1)}}\right)^p-(a q^{(p+1)(q-s)}+bq^{(p+1)(q-s-1)})^p \\
=&(a q^{(p+1)(q-s)})^p+q^{p((p+1)(q-s)-1)}-(a q^{(p+1)(q-s)}+bq^{(p+1)(q-s-1)})^p\\
=&q^{p((p+1)(q-s)-1)}-O(q^{(p+1)(q-s)(p-1)}q^{(p+1)(q-s-1)})\\
=&q^{p((p+1)(q-s)-1)}-O(q^{(p+1)(q-s)p-(p+1)})>0.
\end{align*}
The statement follows from the fact that the power function is monotonically increasing.
\end{proof}

Next, we show that the tour $T$ is indeed $k$-optimal for $q$ large enough. 

\begin{lemma} \label{k optimal eulerian}
There exists a $\bar{q}$ such that for all $q\geq \bar{q}$ the tour $T$ for the constructed instance $I_q$ is $k$-optimal.
\end{lemma}

\begin{figure}[!htb]
\centering
\begin{tikzpicture}[scale=0.32]
\def\yb{3}
\def\xd{10}
\def\xa{8}
\def\yc{7}
\tikzstyle{vertex}=[circle,fill, minimum size=5, inner sep=0]
\tikzstyle{arrow}=[Straight Barb[length=1mm]]

\node[blue, vertex, label=below:$x_i$]  (xi)   at (  0, 0) {};
\node[blue, vertex]  (a)   at (  0, 2) {};
\node[blue, vertex]  (b)   at (  0, 4) {};
\node[blue, vertex]  (c)   at (  0, 6) {};
\node[blue, vertex]  (d)   at (  0, 8) {};
\node[blue, vertex, label=below:$x'_i$] (x'i)  at ( 12, 0) {};
\node[blue, vertex, label=below:$x_j$]  (xj)   at ( 24, 0) {};
\node[blue, vertex, label=below:$x'_j$] (x'j)  at ( 36, 0) {};

\draw[red, line width = 1] (xi) to node[below] {$e_i$} (x'i); 
\draw[gray, dash pattern = on 1mm off 1mm, line width = 0.25] (x'i) -- (xj); 
\draw[blue, line width = 1] (xj) to node[below] {$e_j$} (x'j); 
\draw[line width = 1] (a) -- (b); 
\draw[line width = 1] (c) -- (d); 
\draw[gray, dash pattern = on 1mm off 1mm, line width = 0.25] (xi) -- (a); 
\draw[gray, dash pattern = on 1mm off 1mm, line width = 0.25] (b) -- (c); 
\draw[red, dash pattern = on 0.33mm off 0.33mm, line width = 1] (d) to node[above] {$f'_i$} (12, 16/3); 
\draw[gray, dash pattern = on 1mm off 1mm, line width = 0.25] (12,16/3) -- (24,8/3); 
\draw[blue, dash pattern = on 0.33mm off 0.33mm, line width = 1] (x'j) to node[above] {$f'_j$} (24, 8/3); 

\draw[gray, line width = 0.25] (d) -- (-1, 8); 
\draw[gray, line width = 0.25] (12, 16/3) -- (-1, 16/3); 
\draw[line width = 1]    (-1.75, 24/3) to node[left] {$d_i$} (-1.75, 16/3); 
\draw[line width = 0.25] (-2, 24/3) -- (-1.5, 24/3); 
\draw[line width = 0.25] (-2, 16/3) -- (-1.5, 16/3);

\end{tikzpicture}
  \caption{Sketch for Case 2 in the proof of Lemma~\ref{k optimal eulerian}: The solid edges are the tour edges and the dotted edges are the non-tour edges of $C$. The set $S$ consists of the red and blue solid edges. The red and blue tour edges are assigned to the red and blue dotted parts, respectively.}
  \label{figure alternating walk}
\end{figure}
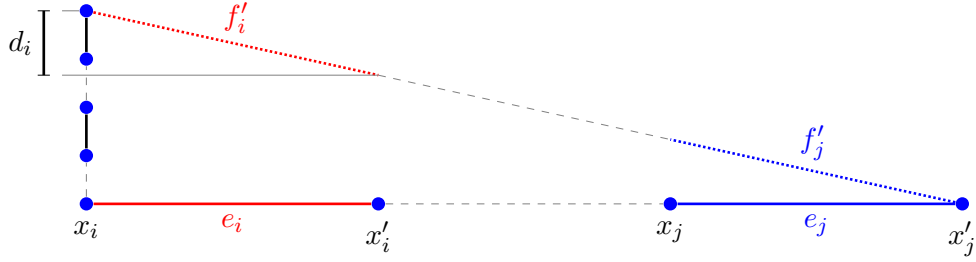

\begin{proof}
Assume that $T$ is not $k$-optimal. Then there is a closed alternating walk $C$ with at most $k$ tour edges and positive gain. We distinguish two cases: \medskip

\noindent Case 1: $C$ visits vertices from at least two layers of $I_q$.\medskip

Assume that $l_s$ is the layer $C$ visits with the smallest index $s$. Moreover, let $C$ contain exactly $a$ tour edges with both endpoints lying on $l_s$. By construction, we can bound the length of the tour edges in $C$ from above by $a q^{(p+1)(q-s)}+(k-a)q^{(p+1)(q-s-1)}$. By assumption, the bounding box of $C$ has side length at least $a q^{(p+1)(q-s)}$ in $x$-direction. Since $C$ contains vertices from at least two layers, its bounding box has side length at least the distance between $l_s$ and $l_{s+1}$ which is $q^{(p+1)(q-s)-1}$ in $y$-direction. By Lemma~\ref{connecting edge} viewing $C$ as a polygon, the edges of $C$ have total length at least $2\sqrt[p]{(a q^{(p+1)(q-s)})^p+q^{((p+1)(q-s)-1)p}}$. Hence, the length of the non-tour edges is at least $2\sqrt[p]{(a q^{(p+1)(q-s)})^p+q^{((p+1)(q-s)-1)p}}-(a q^{(p+1)(q-s)}+(k-a)q^{(p+1)(q-s-1)})$ and by Lemma~\ref{estimate} the total gain of $C$ has to be negative for $q$ large enough, contradiction. \medskip

\noindent Case 2: $C$ visits vertices from at most one layer  of $I_q$.\medskip

Let $S=\{e_1,\dots, e_l\}$ be the set of edges in $C$ with both endpoints lying on the same layer. Moreover, let the endpoints of $e_i$ have the coordinates $(x_i,y_i)$ and $(x_i',y'_i)$ and w.l.o.g.\ assume that $x_i<x_i'$. Since $C$ is a closed walk, it intersects the vertical line $x=h$ for all $h\in\mathbb{R}$ an even number of times. This means that for $x_i<h<x_i'$ the vertical line $x=h$ intersects $C$ in a set of edges $J_h$ with $\lvert J_h\rvert\geq 2$. Furthermore, $J_h$ contains besides of $e_i$ only non-tour edges since otherwise $C$ would visit at last two layers. By construction, $J_h$ contains the same edges for all $x_i<h<x_i'$ since otherwise $C$ would visit more than one layer. Therefore, we can assign every edge $e_i$ in $S$ to an arbitrary non-tour edge $f_i$ in $J_h$ for some $x_i<h<x_i'$. Note that a non-tour edge may be assigned to multiple tour edges in $S$. We mark all edges in $S$ and the corresponding edges they are assigned to.

We claim that the length of the marked non-tour edges is at least as long as the marked tour edges. To see this we assign the length of edges $e_i\in S$ to that part of $f_i$ that the vertical line $x=h$ intersects for values of $h$ satisfying $x_i<h<x_i'$ (Figure~\ref{figure alternating walk}). We call that part of the edge $f'_i$. Note that $f'_i$ and $f'_j$ are disjoint for $i\neq j$ since the interior of $e_i$ and $e_j$ have disjoint $x$-coordinates. Let $d_i$ be the length of the projection of $f'_i$ to the $y$-axis. As $\Vert f'_i\Vert_p = \sqrt[p]{d_i^p+(x'_i-x_i)^p}\geq \sqrt[p]{(x'_i-x_i)^p}=x'_i-x_i$ the length of $f'_i$ is at least as long as that of $e_i$. This proves the claim.

Since we marked one non-tour edge for every tour edge in $S$, there are at least as many unmarked non-tour edges as unmarked tour edges. Note that the length of each of the unmarked non-tour edges is at least~1 and that of each of the unmarked tour edges is exactly~1. Therefore, the total length of the unmarked non-tour edges is larger than or equal to the length of the unmarked tour edges. Combining with the results for the marked edges we see that $C$ cannot have positive gain, contradiction.
\end{proof}

\begin{theorem}
For constant $k\ge 2$ the approximation ratio of the $k$-Opt heuristic for instances whose distances arise from the $p$-norm \blue{with $1\le p < \infty$} is $\Omega\left(\frac{\log(n)}{\log\log(n)}\right)$ where $n$ is the number of vertices.
\label{thm:LowerBoundLp}
\end{theorem}

\begin{proof}
By Lemma~\ref{k optimal eulerian} there is a $q$ such that the tour $T$ for the instance $I_q$ is $k$-optimal. By Lemma~\ref{k-Opt Euc T bound} and~\ref{k-Opt Euc opt bound} the tour $T$ has length at least $q\cdot q^{(p+1)q}$ while the optimal tour $T^*$ for $I_q$ has length at most $14q^{(p+1)q}$. Therefore, the approximation ratio is at least $\frac{q\cdot q^{(p+1)q}}{14q^{(p+1)q}}=\frac{q}{14}$. Recall that $n\in\Theta\left(q^{(p+1)q}\right)$ and hence $q\in\Theta\left(\frac{\log n}{\log \log n}\right)$. 
\end{proof}

In particular, for $p=2$ we get the result for the Euclidean TSP.

\begin{corollary}
For constant $k\ge 2$ the approximation ratio of the $k$-Opt heuristic for Euclidean TSP instances is $\Omega\left(\frac{\log(n)}{\log\log(n)}\right)$ where $n$ is the number of vertices.
\end{corollary}

\section{Higher Dimensional Instances}  
\label{sec:higherDimension}

Most steps of our proof of Theorem~\blue{\ref{thm:main}} not only hold for the 2-dimensional case but also hold in higher dimensions. This is especially true for all results proven in \blue{Sections~\ref{sec:proofidea} and~\ref{sec:arborescence-lemmas}}.
However, there is one step in our proof that we do not know how to extend to higher dimensions. This is the partition of the edge set of the 2-optimal tour into the five subsets $S_1, S'_1, S'_2, S'_2$, and $S_3$. The property that we need is that within each of these five sets no two edges cross. For our proof it is not important that we have five sets. Any constant number of sets would do it. Thus the question for higher dimensions is: 

\begin{quote}
Let $V$ be a $d$-dimensional Euclidean TSP instance, $T$ an optimum tour for $V$ and $S$ a 2-optimal tour for $V$. Embed $V$ into the plane such that $T$ is a plane graph. 
Let $V'\subseteq \mathbb{R}^2$ be the set of points obtained by adding to the embedding of $V$ all
crossings between pairs of edges in $T$ and $S$ and $S'$ be the tour induced by $V'$ for $S$ (as described in Section~\ref{sec:uncrossing}). 
Is it possible to partition the edge set of $S'$ into a constant number of subsets such that each of these subsets together with $T$ forms a planar graph?      
\end{quote}

Unfortunately, the answer to the above question is ''no`` for each plane embedding of $T$. In the following we will construct for each even number $k$ a 3-dimensional Euclidean TSP instance $I_k$ that does not have a plane embedding with the above property. Our instance contains $4k$ points labeled $A_1, \ldots, A_k$, $B_1, \ldots, B_k$, $C_1, \ldots, C_k$, and $D_1, \ldots, D_k$. For each $i$ with $0\le i\le k$ the coordinates for these points are defined as follows: 
\begin{eqnarray*}
A_i & := & (i, 0, 0) \\
B_i & := & (i, 1, 0) \\
C_i & := & (i, 1/2, \sqrt 3 /2) \\
D_i & := & (i, 3/2, \sqrt 3 /2) 
\end{eqnarray*}

This definition implies that for all $i$ with $0\le i\le k$ the segments $A_iB_i$, $A_iC_i$, $B_iC_i$, $B_iD_i$, and
$C_iD_i$ have length~1. Moreover, the segments $A_iA_{i+1}$, $B_iB_{i+1}$, $C_iC_{i+1}$, and $D_iD_{i+1}$ have length~1 for
$0\le i < k$. The segments between any other two points have length strictly larger than~1. Figure~\ref{fig:Ik} shows as an example
the instance $I_8$ and all segments of length~1 within this instance.

\newcommand{\myGlobalTransformation}[2]
{
    \pgftransformcm{1}{0}{0.4}{0.5}{\pgfpoint{#1cm}{#2cm}}
}

\newcommand{\gridThreeD}[3]
{
    \begin{scope}
        \myGlobalTransformation{#1}{#2};
        \fill[white,fill opacity=0.6] (0,0) rectangle (9,3);
        \draw [#3,step=1.25cm] grid (9,3);
    \end{scope}
}

\newcommand{\drawLinewithBG}[3]
{
    \draw[white, line width=3pt,opacity=1.0]  (#2) -- (#3);
    \draw[#1, very thick] (#2) -- (#3);
}

\newcommand{\drawLine}[3]
{
    \draw[#1, very thick] (#2) -- (#3);
}

\begin{figure}[!htb]
\centering

\begin{tikzpicture}[x=1.25cm, y=1.25cm, z=1.25cm]

\def\sqrtthree{1.732050808}
      
\gridThreeD{0}{0}{black!50};

\tikzstyle{vertex}=[circle,fill, minimum size=5, inner sep=0]
       
\foreach \x in {1,...,8} {
  \node[blue, vertex, label=  -40:{\hspace*{-3mm}\footnotesize $A_\x$}]  (A\x)  at (\x + 1 * 0.4, 1 * 0.5) {};
  \node[blue, vertex, label=  -40:{\hspace*{-3mm}\footnotesize $B_\x$}]  (B\x)  at (\x + 2 * 0.4, 2 * 0.5) {};
  \node[blue, vertex, label=  -5:{\hspace*{-13mm}\rlap{\footnotesize $C_\x$}}]  (C\x)  at (\x + 1 * 0.4 + 0.5 * 0.4, 1 * 0.5 + \sqrtthree * 0.5 * 0.8 + 0.5 * 0.5) {};
  \node[blue, vertex,  label=  -5:{\hspace*{-2mm}\rlap{\footnotesize $D_\x$}}]  (D\x)  at (\x + 2 * 0.4 + 0.5 * 0.4, 2 * 0.5 + \sqrtthree * 0.5 * 0.8 + 0.5 * 0.5) {};
}  

\drawLine{black}{A1}{A8}
\drawLine{black}{B1}{B8}
       
\foreach \x in {1,...,8} {
  \drawLine{black}{A\x}{B\x}
  \drawLinewithBG{black}{C\x}{D\x}
  \drawLinewithBG{black}{A\x}{C\x}
  \drawLinewithBG{black}{B\x}{D\x}
  \drawLine{black}{B\x}{C\x}
}

\drawLinewithBG{black}{C1}{C8}
\drawLine{black}{D1}{D8}

\foreach \x in {1,...,8} {
  \node[blue, vertex]  (A\x)  at (\x + 1 * 0.4, 1 * 0.5) {};
  \node[blue, vertex]  (B\x)  at (\x + 2 * 0.4, 2 * 0.5) {};
  \node[blue, vertex]  (C\x)  at (\x + 1 * 0.4 + 0.5 * 0.4, 1 * 0.5 + \sqrtthree * 0.5 * 0.8 + 0.5 * 0.5) {};
  \node[blue, vertex]  (D\x)  at (\x + 2 * 0.4 + 0.5 * 0.4, 2 * 0.5 + \sqrtthree * 0.5 * 0.8 + 0.5 * 0.5) {};
}  
      
\end{tikzpicture}
\caption{The points of the 3-dimensional  instance $I_8$ and the graph on these points induced by all segments of length $1$.}
\label{fig:Ik}
\end{figure}
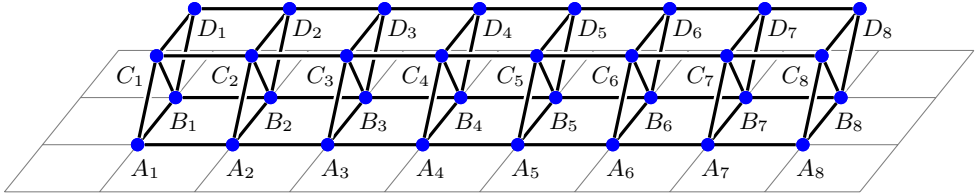

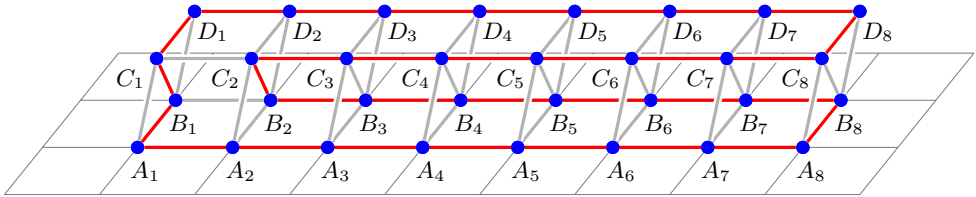
\begin{figure}[!htb]
\centering
\begin{tikzpicture}[x=1.25cm, y=1.25cm, z=1.25cm]

\def\sqrtthree{1.732050808}
      
\gridThreeD{0}{0}{black!50};

\tikzstyle{vertex}=[circle,fill, minimum size=5, inner sep=0]
       
\foreach \x in {1,...,8} {
  \node[blue, vertex, label=  -40:{\hspace*{-3mm}\footnotesize $A_\x$}]  (A\x)  at (\x + 1 * 0.4, 1 * 0.5) {};
  \node[blue, vertex, label=  -40:{\hspace*{-3mm}\footnotesize $B_\x$}]  (B\x)  at (\x + 2 * 0.4, 2 * 0.5) {};
  \node[blue, vertex, label=  -5:{\hspace*{-13mm}\rlap{\footnotesize $C_\x$}}]  (C\x)  at (\x + 1 * 0.4 + 0.5 * 0.4, 1 * 0.5 + \sqrtthree * 0.5 * 0.8 + 0.5 * 0.5) {};
  \node[blue, vertex,  label=  -5:{\hspace*{-2mm}\rlap{\footnotesize $D_\x$}}]  (D\x)  at (\x + 2 * 0.4 + 0.5 * 0.4, 2 * 0.5 + \sqrtthree * 0.5 * 0.8 + 0.5 * 0.5) {};
}  

\drawLine{black!30}{A1}{A8}
\drawLine{black!30}{B1}{B8}
\drawLine{red}{B2}{B8}
\drawLine{red}{A1}{A8}

\foreach \x in {1,...,8} {
  \drawLine{black!30}{A\x}{B\x}
  \drawLinewithBG{black!30}{C\x}{D\x}
  \drawLinewithBG{black!30}{A\x}{C\x}
  \drawLinewithBG{black!30}{B\x}{D\x}
  \drawLine{black!30}{B\x}{C\x}
}  

  \drawLine{red}{A1}{B1}
  \drawLine{red}{C1}{B1}
  \drawLinewithBG{red}{C1}{D1}
  \drawLinewithBG{red}{A8}{B8}
  \drawLine{red}{B2}{C2}
  \drawLinewithBG{red}{C8}{D8}

\drawLinewithBG{black!30}{C1}{C8}
\drawLine{red}{D1}{D8}
\drawLinewithBG{red}{C2}{C8}

\foreach \x in {1,...,8} {
  \node[blue, vertex]  (A\x)  at (\x + 1 * 0.4, 1 * 0.5) {};
  \node[blue, vertex]  (B\x)  at (\x + 2 * 0.4, 2 * 0.5) {};
  \node[blue, vertex]  (C\x)  at (\x + 1 * 0.4 + 0.5 * 0.4, 1 * 0.5 + \sqrtthree * 0.5 * 0.8 + 0.5 * 0.5) {};
  \node[blue, vertex]  (D\x)  at (\x + 2 * 0.4 + 0.5 * 0.4, 2 * 0.5 + \sqrtthree * 0.5 * 0.8 + 0.5 * 0.5) {};
}  
      
\end{tikzpicture}
\caption{The optimum tour $T$ in the instance $I_8$.} 
\label{fig:OptTour}
\end{figure}

Within the instance $I_k$ we can define two optimal (and therefore also 2-optimal) tours using only edges of length~1 as follows. The tour $T$ contains the following edges (see Figure~\ref{fig:OptTour}):

\begin{eqnarray*}
\bigcup_{i=1}^{k-1}\left\{\{A_i, A_{i+1}\},\{D_i, D_{i+1}\}\right\} \cup
\bigcup_{i=2}^{k-1}\left\{\{B_i, B_{i+1}\},\{C_i, C_{i+1}\}\right\} \\
\cup \left\{ \{A_1,B_1\},\{B_1,C_1\},\{C_1,D_1\},\{B_2,C_2\},\{A_k,B_k\},\{C_k,D_k\}
\right\}
\end{eqnarray*}

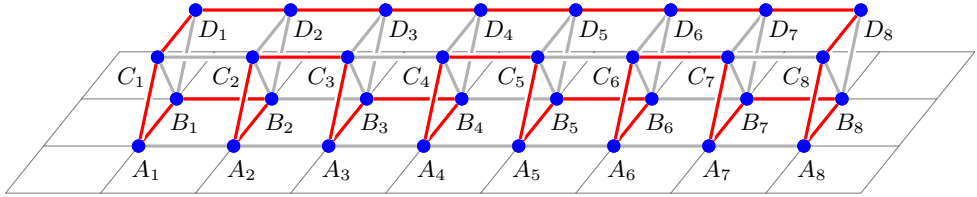
\begin{figure}[!htb]
\centering
\begin{tikzpicture}[x=1.25cm, y=1.25cm, z=1.25cm]

\def\sqrtthree{1.732050808}
      
\gridThreeD{0}{0}{black!50};

\tikzstyle{vertex}=[circle,fill, minimum size=5, inner sep=0]
       
\foreach \x in {1,...,8} {
  \node[blue, vertex, label=  -40:{\hspace*{-3mm}\footnotesize $A_\x$}]  (A\x)  at (\x + 1 * 0.4, 1 * 0.5) {};
  \node[blue, vertex, label=  -40:{\hspace*{-3mm}\footnotesize $B_\x$}]  (B\x)  at (\x + 2 * 0.4, 2 * 0.5) {};
  \node[blue, vertex, label=  -5:{\hspace*{-13mm}\rlap{\footnotesize $C_\x$}}]  (C\x)  at (\x + 1 * 0.4 + 0.5 * 0.4, 1 * 0.5 + \sqrtthree * 0.5 * 0.8 + 0.5 * 0.5) {};
  \node[blue, vertex,  label=  -5:{\hspace*{-2mm}\rlap{\footnotesize $D_\x$}}]  (D\x)  at (\x + 2 * 0.4 + 0.5 * 0.4, 2 * 0.5 + \sqrtthree * 0.5 * 0.8 + 0.5 * 0.5) {};
}  

\drawLine{black!30}{A1}{A8}
\drawLine{black!30}{B1}{B8}
\drawLine{red}{B1}{B2}
\drawLine{red}{B3}{B4}
\drawLine{red}{B5}{B6}
\drawLine{red}{B7}{B8}

\foreach \x in {1,...,8} {
  \drawLine{red}{A\x}{B\x}
  \drawLinewithBG{black!30}{C\x}{D\x}
  \drawLinewithBG{red}{A\x}{C\x}
  \drawLinewithBG{black!30}{B\x}{D\x}
  \drawLine{black!30}{B\x}{C\x}
}  

\drawLinewithBG{red}{C1}{A1}
\drawLinewithBG{red}{C1}{D1}
\drawLinewithBG{red}{C8}{D8}

\drawLinewithBG{black!30}{C1}{C8}

\drawLinewithBG{red}{C3}{C2}
\drawLinewithBG{red}{C5}{C4}
\drawLinewithBG{red}{C7}{C6}
  
\drawLine{red}{D1}{D8}

\foreach \x in {1,...,8} {
  \node[blue, vertex]  (A\x)  at (\x + 1 * 0.4, 1 * 0.5) {};
  \node[blue, vertex]  (B\x)  at (\x + 2 * 0.4, 2 * 0.5) {};
  \node[blue, vertex]  (C\x)  at (\x + 1 * 0.4 + 0.5 * 0.4, 1 * 0.5 + \sqrtthree * 0.5 * 0.8 + 0.5 * 0.5) {};
  \node[blue, vertex]  (D\x)  at (\x + 2 * 0.4 + 0.5 * 0.4, 2 * 0.5 + \sqrtthree * 0.5 * 0.8 + 0.5 * 0.5) {};
}  
      
\end{tikzpicture}
\caption{The 2-optimal tour $S$ in the instance $I_8$.} 
\label{fig:2OptTour}
\end{figure}

The tour $S$ contains the following edges (see Figure~\ref{fig:2OptTour}):
\begin{eqnarray*}
\left\{ \{C_1,D_1\}, \{C_k,D_k\}\right\} \cup
\bigcup_{i=1}^{k-1}\left\{\{D_i, D_{i+1}\}\right\} \cup
\bigcup_{i=1}^{k}  \left\{\{A_i, B_i\}, \{A_i, C_i\}\right\} \cup
\bigcup_{i=1}^{k/2}  \left\{\{B_{2i-1}, B_{2i}\}\right\} 
\end{eqnarray*}

Now embed the points of $I_k$ into the plane such that $T$ is a plane graph. 
Then there will be a linear number of pairwise intersections between the edges $\{A_i, C_i\}$ in $S$ for $0\le i\le k$. 
Thus it is not possible to partition the edges of $S$
into a constant number of subsets such that each of them has no pair of intersecting edges.

\section{Extension to arbitrary \texorpdfstring{$p$}{p}-norms}
\label{sec:LpNorms}

Our proof of \blue{Theorem~\ref{thm:combinedresult}} is mostly independent of the Euclidean norm. 
\blue{Theorem~\ref{thm:LowerBoundLp} shows that the lower bound holds for arbitrary $p$-norms (with $1\le p < \infty$). 
All} the results
presented in Section~\ref{sec:arborescence-lemmas} work for arbitrary metrics. The only crucial part of
our argument \blue{for the lower bound} is our use of planarity and for this we need Lemma~\ref{lemma:nocrossing}. This lemma holds for arbitrary $p$-norms
\blue{with $p > 1$} and therefore we can replace the $2$-norm in our proof by arbitrary $p$-norms \blue{with $1 < p < \infty$}. 

\blue{The case $p=1$ needs some special care as optimal tours may contain crossings in this case. 
A result of van Leeuwen and Schoone~\cite{LS1982} also applies to the 1-norm: After applying $O(n^3)$ 2-moves,
each removing a crossing, to an arbitrary tour we get a tour without crossings. We apply this procedure to an optimal tour and end up with an optimal tour without crossings since every 2-move that removes a crossing does not increase the length of the tour.}

 \blue{For a 2-optimal tour we cannot apply the above argument in the case $p=1$. 
By performing a 2-move that does not change the length of a 2-optimal tour it may happen that the new tour is no longer 2-optimal. 
We use a different argument to handle crossings in 2-optimal tours in the case $p=1$.
Assume that two edges $e$ and $f$ intersect. Let $e_x$, $e_y$, $f_x$, and $f_y$ be the width and height of the smallest axis-parallel rectangle
containing $e$ and $f$, respectively. 
By a case distinction we see that a 2-move replacing $e$ and $f$ shortens the tour by $2 e_x$, $2 e_y$, $2 f_x$, or $2 f_y$ 
depending on the position of $e$ and $f$. If both edges are neither parallel to the $x$-axis nor parallel to the $y$-axis, all these terms 
are strictly positive and hence there is a 2-move that shortens the tour which is a contradiction. 
Thus, either $e$ or $f$ is parallel to the $x$- or $y$-axis. Now we can divide the edges of
the 2-optimal tour into three sets: those parallel to the $x$-axis, those parallel to the $y$-axis and the remaining edges. 
The edges in each set do not intersect each other and we can bound the total length of the edges in each set separately 
by applying the approach of Section~\ref{sec:edge-partition} to each of these sets. By this we loose at most a factor of~3. }

Thus we get as our final main result:
\begin{theorem} For constant $k \ge 2$ the 
approximation ratio of the $k$-Opt heuristic for $n$ points in the plane with distances measured by the $p$-norm \blue{with $1\le p < \infty$} is $\Theta(\log n / \log \log n)$.
\label{thm:LpNormResult}
\end{theorem}

\bibliographystyle{siamplain}
\bibliography{EuclideankOPT_arxive_revised} 

\begin{thebibliography}{10}

\bibitem{Aro1998}
{\sc S.~Arora}, {\em Polynomial time approximation schemes for {E}uclidean
  {T}raveling {S}alesman and other geometric problems}, Journal of the ACM, 45
  (1998), pp.~753--782.

\bibitem{Ben1992}
{\sc J.~J. Bentley}, {\em Fast algorithms for geometric {T}raveling {S}alesman
  {P}roblems}, ORSA Journal on Computing, 4 (1992), pp.~387--411.

\bibitem{CKT1999}
{\sc B.~Chandra, H.~Karloff, and C.~Tovey}, {\em New results on the old $k$-opt
  algorithm for the {T}raveling {S}alesman {P}roblem}, SIAM J. Comput., 28
  (1999), pp.~1998--2029.

\bibitem{Chr2022b}
{\sc N.~Christofides}, {\em Worst-case analysis of a new heuristic for the
  travelling salesman problem}, Operations Research Forum, 3 (2022), p.~Article
  20.

\bibitem{ERV2014}
{\sc M.~Englert, H.~R\"oglin, and B.~V\"ocking}, {\em Worst case and
  probabilistic analysis of the 2-opt algorithm for the {TSP}}, Algorithmica,
  68 (2014), pp.~190--264.

\bibitem{Flo1956}
{\sc M.~M. Flood}, {\em The traveling-salesman problem}, Operations Research, 4
  (1956), pp.~61--75.

\bibitem{GJ1979}
{\sc M.~R. Garey and D.~S. Johnson}, {\em Computers and Intractability. A Guide
  to the Theory of {NP}-Completeness}, W. H. Freeman and Company, 1979.

\bibitem{HZZ2020}
{\sc S.~Hougardy, F.~Zaiser, and X.~Zhong}, {\em The approximation ratio of the
  2-{O}pt~{H}euristic for the metric {T}raveling {S}alesman {P}roblem},
  Operations Research Letters, 48 (2020), pp.~401--404.

\bibitem{KKO2022}
{\sc A.~R. Karlin, N.~Klein, and S.~O. Gharan}, {\em A (slightly) improved
  deterministic approximation algorithm for metric {TSP}}.
\newblock arXiv:2212.06296v1 [cs.DS], Dec. 2022.

\bibitem{KMSV1998}
{\sc S.~Khanna, R.~Motwani, M.~Sudan, and U.~Vazirani}, {\em On syntactic
  versus computational views of approximability}, SIAM J. Comput., 28 (1998),
  pp.~164--191.

\bibitem{Mit1999}
{\sc J.~S.~B. Mitchell}, {\em Guillotine subdivisions approximate polygonal
  subdivisions: A simple polynomial-time approximation scheme for geometric
  {TSP}, $k$-{MST}, and related problems}, SIAM Journal on Computing, 28
  (1999), pp.~1298--1309.

\bibitem{Pap1977}
{\sc C.~H. Papadimitriou}, {\em The {E}uclidean {T}raveling {S}alesman
  {P}roblem is {NP}-complete}, Theoretical Computer Science, 4 (1977),
  pp.~237--244.

\bibitem{Rei1994}
{\sc G.~Reinelt}, {\em The Traveling Salesman. Computational Solutions for TSP
  Applications}, Springer, 1994.

\bibitem{SG1976}
{\sc S.~Sahni and T.~Gonzalez}, {\em P-complete approximation problems},
  Journal of the Association for Computing Machinery, 23 (1976), pp.~555--565.

\bibitem{Ser1978}
{\sc A.~Serdyukov}, {\em O nekotorykh ekstremal'nykh obkhodakh v grafakh},
  Upravlyaemye sistemy, 17 (1978), pp.~76--79.

\bibitem{LS1982}
{\sc J.~van Leeuwen and A.~A. Schoone}, {\em Untangling a travelling salesman
  tour in the plane}, in Proceedings of the 7th Conference on Graphtheoretic
  Concepts in Computer Science (WG81), J.~R. M\"uhlbacher, ed., Carl Hanser
  Verlag, 1982, pp.~88--98.

\bibitem{Zho2020}
{\sc X.~Zhong}, {\em Approximation Algorithms for the {T}raveling {S}alesman
  {P}roblem}, PhD thesis, Research Institute for Discrete Mathematics,
  University of Bonn, 2020.

\bibitem{Zho2020b}
{\sc X.~Zhong}, {\em On the approximation ratio of the $k$-opt and
  {L}in-{K}ernighan algorithm for metric and graph {TSP}}, in 28th Annual
  European Symposium on Algorithms (ESA 2020), F.~Grandoni, G.~Herman, and
  P.~Sanders, eds., vol.~173 of Leibniz International Proceedings in
  Informatics (LIPIcs), 2020, pp.~83:1--83:13,
  \url{https://doi.org/10.4230/LIPIcs.ESA.2020.83},
  \url{https://drops.dagstuhl.de/opus/volltexte/2020/12949}.

\bibitem{Zho2021}
{\sc X.~Zhong}, {\em On the approximation ratio of the 3-opt algorithm for the
  (1,2)-{TSP}}, Operations Research Letters, 49 (2021), pp.~515--521,
  \url{https://doi.org/https://doi.org/10.1016/j.orl.2021.05.012},
  \url{https://www.sciencedirect.com/science/article/pii/S0167637721000808}.

\end{thebibliography}

\end{document}